\pdfoutput=1
\documentclass[10pt,letterpaper]{article}
\usepackage[top=0.85in,left=2.75in,footskip=0.75in]{geometry}

\usepackage{amsmath,amssymb}

\usepackage{changepage}

\usepackage[utf8x]{inputenc}

\usepackage{textcomp,marvosym}

\usepackage{cite}

\usepackage{nameref,hyperref}

\usepackage[right]{lineno}

\usepackage{microtype}
\DisableLigatures[f]{encoding = *, family = * }

\usepackage[table]{xcolor}

\usepackage{array}

\newcolumntype{+}{!{\vrule width 2pt}}

\newlength\savedwidth

\usepackage{setspace} 
\raggedright
\usepackage[aboveskip=1pt,labelfont=bf,labelsep=period,justification=raggedright,singlelinecheck=off]{caption}

\makeatletter
\renewcommand{\@biblabel}[1]{\quad#1.}
\makeatother

\date{}

\usepackage{lastpage,fancyhdr,graphicx}
\usepackage{epstopdf}
\fancyheadoffset[L]{2.25in}
\fancyfootoffset[L]{2.25in}
\begin{document}
\vspace*{0.2in}

% Title must be 250 characters or less.
% Please capitalize all terms in the title except conjunctions, prepositions, and articles.
% oldtitle; Filament Recycling and Sustained Contractile Flows in an Actomyosin Cortex
\begin{flushleft}
{\Large
\textbf\newline{Filament turnover is essential for continuous long range contractile flow in a model actomyosin cortex.}
}
\newline
% Insert author names, affiliations and corresponding author email (do not include titles, positions, or degrees).
\\
William M McFadden\textsuperscript{1},
Patrick M McCall\textsuperscript{2},
Edwin M Munro\textsuperscript{3,*}
\\
\bigskip
\bf{1} Biophysical Sciences Program, University of Chicago, Chicago, IL, USA
\\
\bf{2} Department of Physics, University of Chicago, Chicago, IL, USA
\\
\bf{3} Department of Molecular Genetics and Cell Biology, University of Chicago, Chicago, IL, USA
\\
\bigskip

% Insert additional author notes using the symbols described below. Insert symbol callouts after author names as necessary.
% 
% Remove or comment out the author notes below if they aren't used.
%
% Primary Equal Contribution Note
%\Yinyang These authors contributed equally to this work.

% Additional Equal Contribution Note
% Also use this double-dagger symbol for special authorship notes, such as senior authorship.
%\ddag These authors also contributed equally to this work.

% Current address notes
%\textcurrency a Insert current address of first author with an address update
% \textcurrency b Insert current address of second author with an address update
% \textcurrency c Insert current address of third author with an address update

% Deceased author note
%\dag Deceased

% Group/Consortium Author Note
%\textpilcrow Membership list can be found in the Acknowledgments section.

% Use the asterisk to denote corresponding authorship and provide email address in note below.
* emunro@uchicago.edu

\end{flushleft}
% Please keep the abstract below 300 words
\section*{Abstract}
Actomyosin-based cortical flow is a fundamental engine for cellular morphogenesis.  Cortical flows are generated by cross-linked networks of actin filaments and myosin motors, in which active stress produced by motor activity is opposed by passive resistance to network deformation.  Continuous flow requires local remodeling through crosslink unbinding and and/or filament disassembly. But how local remodeling tunes stress production and dissipation, and how this in turn shapes long range flow, remains poorly understood. Here, we introduce a computational model for a cross-linked networks with active motors based on minimal requirements for production and dissipation of contractile stress, namely asymmetric filament compliance, spatial heterogeneity of motor activity, reversible cross-links and filament turnover.  We characterize how the production and dissipation of network stress depend, individually, on cross-link dynamics and filament turnover, and how these dependencies combine to determine overall rates of cortical flow. Our analysis predicts that filament turnover is required to maintain active stress against external resistance and steady state flow in response to external stress. Steady state stress increases with filament lifetime up to a characteristic time $\tau_{m}$, then decreases with lifetime above $\tau_{m}$.   Effective viscosity increases with filament lifetime up to a characteristic time $\tau_c$, and then becomes independent of filament lifetime and sharply dependent on crosslink dynamics.  These individual dependencies of active stress and effective viscosity define multiple regimes of steady state flow.  In particular our model predicts the existence of a regime, when filament lifetimes are shorter than both $\tau_c$ and $\tau_{m}$, in which dependencies of effective viscosity and steady state stress cancel one another, such that flow speed is insensitive to filament turnover, and shows simple dependence on motor activity and crosslink dynamics.  These results provide a framework for understanding how animal cells tune cortical flow through local control of network remodeling.

% Please keep the Author Summary between 150 and 200 words
% Use first person. PLOS ONE authors please skip this step. 
% Author Summary not valid for PLOS ONE submissions.   
\section*{Author Summary}
In this paper, we develop and analyze a minimal model for a 2D network of cross-linked actin filaments and myosin motors, representing the cortical cytoskeleton of eukaryotic cells.  We implement coarse-grained representations of force production by myosin motors and stress dissipation through an effective cross-link friction and filament turnover. We use this model to characterize how the sustained production of active stress, and the steady dissipation of elastic stress, depend individually on motor activity,  effective cross-link friction and filament turnover. Then we combine these results to gain insights into how microscopic network parameters control steady state flow produced by asymmetric distributions of motor activity. Our results provide a framework for understanding how local modulation of microscopic interactions within contractile networks control macroscopic quantities like active stress and effective viscosity to control cortical deformation and flow at cellular scales.  

%\linenumbers

\section*{Introduction}

\paragraph{}  Cortical flow is a fundamental and ubiquitous form of cellular deformation that underlies cell polarization, cell division, cell crawling and multicellular tissue morphogenesis\cite{cellmech_flows3,cellmech_flows2,Benink:2000aa,Wilson:2010aa,Rauzi2010,Munro2004413}. These flows originate within a thin layer of cross-linked actin filaments and myosin motors, called the actomyosin cortex, that lies just beneath the plasma membrane \cite{Salbreux2012536}. Local forces produced by bipolar myosin filaments are integrated within cross-linked networks to build macroscopic contractile stress\cite{Murrell:2015aa,Bendix20083126,Janson1005}.  At the same time, cross-linked networks resist deformation and this resistance must be dissipated by network remodeling to allow macroscopic deformation and flow.  How force production and dissipation depend on motor activity and network remodeling remains poorly understood.

\paragraph{}  One successful approach to modeling cortical flow has relied on coarse-grained phenomenological descriptions of actomyosin networks as active fluids, whose motions are driven by gradients of active contractile stress and opposed by an effectively viscous resistance\cite{cellmech_flows}.  In these models, spatial variation in active stress is typically assumed to reflect spatial variation in motor activity and force transmission\cite{PhysRevLett.106.028103}, while viscous resistance is assumed to reflect the internal dissipation of elastic resistance due to local remodeling of filaments and/or cross-links\cite{Salbreux2012536, De-La-Cruz:2015aa}. Models combining an active fluid description with simple kinetics for network assembly and disassembly, can successfully reproduce the spatiotemporal dynamics of cortical flow observed during polarization \cite{cellmech_flows}, cell division \cite{Turlier2014114,PhysRevLett.103.058102}, cell motility \cite{Keren:2009aa,RevModPhys.85.1143} and tissue morphogenesis \cite{Behrndt257}. However, it remains a challenge to connect this coarse-grained description of cortical flow to the microscopic origins of force generation and dissipation within cross-linked actomyosin networks.  

\paragraph{} Studies in living cells reveal fluid-like stress relaxation on timescales of 10-100s \cite{cellmech_flows,cellmech_flows2,cellmech_flows3,rheo_fluid,rheo_fluid2,cell_rheo_exp}, which is thought to arise through a combination of cross link unbinding and actin filament turnover \cite{De-La-Cruz:2015aa,De-La-Cruz:2009aa,Salbreux2012536}.  Theoretical \cite{theo_crosslinkslip1,theo_crosslinkslip2} and computational \cite{model_taeyoon,rheo_crosslinkslip2,theo_crosslinkslip3} studies reveal that cross-link unbinding can endow actin networks with complex time-dependent viscoelasticity. However, while cross-link unbinding is sufficient for viscous relaxation (creep) on very long timescales {\em in vitro}, it is unlikely to account for the rapid cortical deformation and flow observed in living cells \cite{rheo_crosslinksmatter,rheo_crosslinkslip1,rheo_crosslinkslip2,rheo_crosslinkslip3,rheo_nonaffine}.  Experimental studies in living cells reveal rapid turnover of cortical actin filaments on timescales comparable to stress relaxation (10-100s) \cite{Robin:2014aa,Fritzsche:2013aa,Fritzschee1501337,Carlsson:2010aa,Lai:2008aa}.  Perturbing turnover can lead to changes in cortical mechanics and in the rates and patterns of cortical flow\cite{Van-Goor:2012aa,Fritzschee1501337}.  However, the specific contributions of actin turnover to stress relaxation and how these depend on network architecture remain unclear.   

\paragraph{} Recent work has also begun to reveal mechanisms for active stress generation in disordered actomyosin networks. Theoretical studies suggest that spatial heterogeneity in motor activity along individual filaments, and asymmetrical filament compliance (stiffer in extension than in compression), are sufficient for macroscopic contraction \cite{1367-2630-14-3-033037,PhysRevX.4.041002}, although other routes to contractility may also exist \cite{PhysRevX.4.041002}.  Local interactions among actin filaments and myosin motors are sufficient to drive macroscopic contraction of disordered networks {\em in vitro} \cite{rheo_2D1}, and the kinematics of contraction observed in these studies support a mechanism based on asymmetrical filament compliance and filament buckling.  However, in these studies, the filaments were preassembled and network contraction was transient, because of irreversible network collapse\cite{Alvarado:2013aa}, or buildup of elastic resistance\cite{Murrell15062014}, or because network rearrangements (polarity sorting) dissipate the potential to generate contractile force \cite{Ennomani2016616,Reymann1310,Ndlec:1997aa,Surrey1167}. This suggests that network turnover may play essential role(s) in allowing sustained production of contractile force. Recent theoretical and modeling studies have begun to explore how this might work \cite{2015arXiv150706182H,Mak:2016aa,10.1371/journal.pone.0000696}, and to explore dynamic behaviors that can emerge when contractile material undergoes turnover \cite{PhysRevLett.103.058102,PhysRevLett.113.148102}. However, it remains a challenge to understand how force production and dissipation depend individually on the local interplay of network architecture, motor activity and filament turnover, and how these dependencies combine to mediate tunable control of long range cortical flow. 

\paragraph{}  Here, we construct and analyse a simple computational model that bridges between the microscopic description of cross-linked actomyosin networks and the coarse grained description of an active fluid.  We represent actin filaments as simple springs with asymmetric compliance; we represent dynamic binding/unbinding of elastic cross-links as molecular friction \cite{theo_friction,theo_frictionSam,theo_molefric} at filament crossover points; we represent motor activity as force coupling on a subset of filament cross-over points with a simple linear force/velocity relationship \cite{theo_frictionShila}.  Finally, we model filament turnover by allowing entire filaments to appear with a fixed probability per unit area and disappear with fixed probabilities per unit time. We use this model first to characterize the passive response of a cross-linked network to externally applied stress, then the buildup and maintenance of active stress against an external resistance, and finally the steady state flows produced by an asymmetric distribution of active motors in which active stress and passive resistance are dynamically balanced across the network.  Our results reveal how network remodeling can tune cortical flow through simultaneous effects on active force generation and passive resistance to network deformation.

% You may title this section ``Methods'' or ``Models". 
% ``Models'' is not a valid title for PLoS ONE authors. However, PLoS ONE
% authors may use ``Analysis'' 
\section*{Models}

Our goal was to construct a minimal model that is detailed enough to capture essential microscopic features of cross-linked actomyosin networks (actin filaments with asymmetric compliance, dynamic cross-links, active motors and and continuous filament turnover), but simple enough to explore, systematically, how these microscopic features control macroscopic deformation and flow. We focus on 2D networks because they capture a reasonable approximation of the quasi-2D cortical actomyosin networks that govern flow and deformation in many eukaryotic cells\cite{cellmech_flows, salbreuxbphs}, or the quasi-2D networks studied recently {\em in vitro} \cite{rheo_2D1,rheo_2D2}.

\begin{figure}[h!]
	\centering
	\includegraphics[width=\hsize]{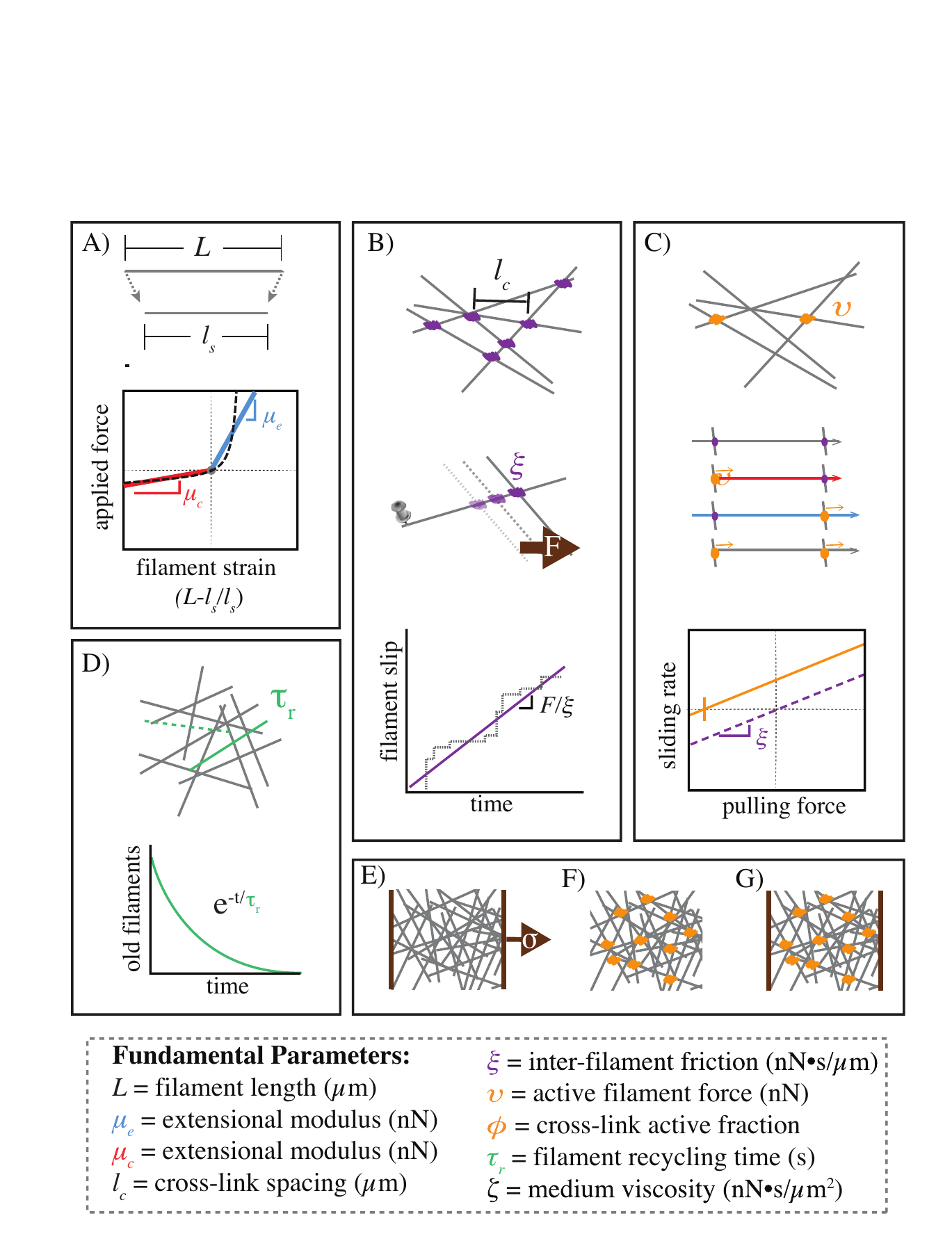}
	\caption{\label{fig:model_overview} Schematic overview of modeling framework and assumptions. \textbf{A)} Filaments are oriented linear springs that are stiffer in extension than in compression. \textbf{B)} Cross-linking occurs at all filament crossings; we represent cross link resistance as an effective drag, proportional to the relative velocity of the overlapping filaments. \textbf{C)} We represent motor activity as a linear force-velocity relationship with a fixed force at zero velocity directed towards a filament's (-) end. We implement spatial heterogeneity by imposing motor activity at a fixed fraction of filament crossover points, resulting in variation in the magnitudes of compressive vs extensile vs translational forces along individual filament segments. \textbf{D)} Whole filaments disappear at a constant rate; new filaments appear with random positions and orientations at the constant rate per unit area, such that entire network refreshes on a characteristic timescale $\tau_r$. \textbf{e-g)} Three different simulation scenarios: \textbf{E)} Passive response to uniaxial stress, \textbf{F)} Free contraction of an active network and \textbf{G)} Isometric contraction against a fixed boundary. }
\end{figure}

Fig. \ref{fig:model_overview} provides a schematic overview of our model's assumptions. We model each filament as an oriented elastic spring with relaxed length $l_s$. The state of a filament is defined by the positions of its endpoints $\mathbf{x_i}$ and $\mathbf{x_{i+1}}$ marking its (-) and (+) ends respectively. The index i enumerates over all endpoints of all filaments. We refer to the filament connecting endpoint i and i+1 as filament i, and we define $\mathbf{\hat{u_i}}$ to be the unit vector oriented along filament i from endpoint i to endpoint i+1.

\subsection*{Asymmetric filament compliance}
We assume (Fig. \ref{fig:model_overview}A) that local deformation of filament  i gives rise to an elastic force:

\begin{equation}
\label{eqn:spring}
\mathbf{F^{\mu}_{i,i+1}} = \mu \gamma_{i}  \mathbf{\hat{u_i}}
\end{equation}

where $ \gamma_{i} = (|\mathbf{x_{i-1}}-\mathbf{x_i}|-l_s)/l_s$ is the strain on filament i, and the elastic modulus  $\mu$ is a composite quantity that represents both filament and cross-linker compliance as in the effective medium theory of Broederz and colleagues \cite{theo_crosslinknonlinear}.  To model asymmetric filament compliance, we set $\mu = \mu_e$ if the strain is positive (extension), and $\mu = \mu_c$ if the strain is negative (compression). The total elastic force on a filament endpoint $\mathbf{i}$ can be written as:

\begin{equation}
\label{eqn:internal}
\mathbf{F^{elas}_i} =  \mathbf{F^{\mu}_{i,i+1}} - \mathbf{F^{\mu}_{i-1,i}} 
\end{equation}

In the limit of highly rigid cross-links and flexible filaments, our model approaches the pure semi-flexible filament models of \cite{theo_hlm,theo_hlm2}. In the opposite limit (nearly rigid filaments and highly flexible cross links), our model approaches that of \cite{theo_crosslinknonlinear} in small strain regimes before any nonlinear cross link stiffening. 

\subsection*{Drag-like coupling between overlapping filaments}
\label{exp_drag}
Previous models represent cross-linkers as elastic connections between pairs of points on neighboring filaments that appear and disappear with either fixed or force-dependent probabilities \cite{model_taeyoon,theo_crosslinknonlinear}.  Here, we introduce a coarse-grained representation of crosslink dynamics by introducing an effective drag force that couples every pair of overlapping filaments, and which represents a molecular friction arising from the time-averaged contributions of many individual transient crosslinks (Fig. \ref{fig:model_overview}B). This coarse-grained approximation has been shown to be adequate in the case of ionic cross-linking of actin\cite{mol_fric,theo_hydroish2}, and has been used to justify simple force-velocity curves for myosin bound filaments in other contexts \cite{theo_frictionShila}. 

To implement coupling through effective drag, for any pair of overlapping filaments j and k, we write the drag force on filament j as:

\begin{equation}
\label{eqn:drag force}
\mathbf{F^{\xi}_{j,k}} = -\xi  (\mathbf{v_{j}}-\mathbf{v_{k}}) 
\end{equation}

where $\xi$ is the drag coefficient and $\mathbf{v_{j}}$, $\mathbf{v_{k}}$ are the average velocities of filaments j and k. We apportion this drag force to the two endpoints ( j, j+1) of filament j as follows: If $\mathbf{x_{j,k}}$ is the position of the filament overlap, then we assign $(1 - \mathbf{\lambda_{j,k}}) \mathbf{F^{\xi}_{j,k}}$ to endpoint j and $\mathbf{\lambda_{j,k}} \mathbf{F^{\xi}_{j,k}}$ to endpoint j+1, where $\mathbf{\lambda_{j,k}} = |\mathbf{x_{j,k}}-\mathbf{x_j}|/|\mathbf{x_{j+1}}-\mathbf{x_j}|$.

The total crosslink coupling force on endpoint i due to overlaps along filament i and i-1 can then be written:

\begin{equation}
\label{eqn: total drag couple}
\mathbf{F^{xl}_{i}} = \sum_j (1 - \mathbf{\lambda_{i,j}}) \mathbf{F^{\xi}_{i,j}} + \sum_k \mathbf{\lambda_{i-1,k}} \mathbf{F^{\xi}_{i-1,k}}
\end{equation}

where the sums are taken over all filaments j and k that overlap with filaments i and i-1 respectively.  

This model assumes a linear relation between the drag force and the velocity difference between attached filaments.   Although non-linearities can arise through force dependent detachment kinetics and/or non-linear force extension of cross-links, we assume here that these non-linear effects are of second or higher order. 

\subsection*{Active coupling for motor driven filament interactions}

To add motor activity at the point of overlap between two filaments j and k ; for each filament in the pair, we impose an additional force of magnitude $\upsilon$, directed towards its (-) end (Fig. \ref{fig:model_overview}C):

\begin{equation}
\label{eqn:directedmotorforce}
\mathbf{F^{\upsilon}_{i}}=-\upsilon \mathbf{\hat{u_i}}
\end{equation}

and we impose an equal and opposite force on its overlapping partner.  We distribute these forces to filament endpoints as described above for crosslink coupling forces.  Thus, the total force on endpoint i due to motor activity can be written as:

\begin{equation}
\label{eqn:active}
\mathbf{F^{motor}_{i}} = \upsilon \sum_j (1 - \mathbf{\lambda_{i,j}}) \left (\mathbf{\hat{u_{i}}} - \mathbf{\hat{u_j}} \right ) q_{i,j}
                        +  \upsilon \sum_k (\mathbf{\lambda_{i-1,k}}) \left (\mathbf{\hat{u_{i-1}}} - \mathbf{\hat{u_k}} \right ) q_{i-1,k} 
\end{equation}

where j and k enumerate over all filaments that overlap with filaments i and i-1 respectively, and $q_{j,k}$ equals 0 or 1 depending on whether there is an ``active'' motor at this location. To model dispersion of motor activity, we set $q_{i,j}=1$  on a randomly selected subset of filament overlaps, such that $\bar{q}=\phi$, where $\bar{q}$ indicates the mean of $q$ (Fig. \ref{fig:model_overview}C).

\subsection*{Equations of motion}

To write the full equation of motion for a network of actively and passively coupled elastic filaments, we assume the low Reynold's number limit in which inertial forces can be neglected, and we equate the sum of all forces acting on each filament endpoint to zero to obtain:

\begin{equation}
\label{eqn:syst3}
0=-l_s\zeta\mathbf{ v_i} -\mathbf{F^{xl}_i}+ \mathbf{F^{elas}_i}+\mathbf{F^{motor}_i} 
\end{equation}

where the first term represents the hydrodynamic drag on the half-filament adjoining endpoint i with respect to motion against the surrounding fluid, and $\zeta$ is the drag coefficient.

\subsection*{2D network formation}

We used a mikado model approach \cite{Unterberger2014} to initialize a minimal network of overlapping unstressed linear filaments in a rectangular 2D domain. We generate individual filaments by laying down straight lines, of length L, with random position and orientation. We define the density using the average distance between cross-links along a filament, $l_c$. A simple geometrical argument can then be used to derive the number of filaments filling a domain as a function of $L$ and $l_c$ \cite{theo_hlm}.  Here, we use the approximation that the number of filaments needed to tile a rectangular domain of size $D_x \times D_y$  is $2D_xD_y/Ll_c$, and that the length density is therefore simply, $2/l_c$. 

\subsection*{Modeling filament turnover}

In living cells, actin filament assembly is governed by multiple factors that control filament nucleation, branching and elongation. Likewise filament disassembly is governed by multiple factors that promote filament severing and monomer dissociation at filament ends. Here, we implement a very simple model for filament turnover in which entire filaments appear with a fixed rate per unit area, $k_{app}$ and disappear at a rate $k_{diss}\rho$, where $\rho$ is a filament density (Fig. \ref{fig:model_overview}D). With this assumption, in the absence of network deformation, the density of filaments will equilibrate to a steady state density, $k_{app}/k_{diss}$, with time constant $\tau_r = 1/k_{diss}$.   In deforming networks, the density will be set by a competition between strain thinning ($\gamma>0$) or thickening ($\gamma<0$), and density equilibration via turnover. To implement this model, at fixed time intervals $\tau_s < 0.01\cdot\tau_r$ (i.e. 1\% of the equilibration time), we selected a fraction, $\tau_s/\tau_r$, of existing filaments (i.e. less than 1\% of the total filaments) for degradation. We then generated a fixed number of new unstrained filaments $k_{app}\tau_sD_xD_y$ at random positions and orientations within the original domain.   We refer to $k_{diss}=1/\tau_r$ as the turnover rate, and to $\tau_r$ as the turnover time.

\subsection*{Simulation methods}

Further details regarding our simulation approach and references to our code can be found in the Supplementary Information (\nameref{S1_Text} A.1). Briefly, equations 1-7 define a coupled system of ordinary differential equations that can be written in the form:

\begin{equation}
\mathbf{A \cdot \dot x} = \mathbf{f(x)}
\end{equation}

where $\mathbf{x}$ is a vector of filament endpoint positions, $\mathbf{\dot{x}}$ the endpoint velocities, $\mathbf{A }$ is a matrix with constant coefficients that represent crosslink coupling forces between overlapping filaments, and $\mathbf{f(x)}$ represents the active (motor) and elastic forces on filament endpoints. We smoothed all filament interactions, force fields, and constraints linearly over small regions such that the equations contained no sharp discontinuities. We numerically integrate this system of equations to find the time evolution of the positions of all filament endpoints. We generate a network of filaments with random positions and orientations as described above within a domain of size $D_x$ by $D_y$.  For all simulations, we imposed periodic boundaries in the y-dimension. To impose an extensional stress, we constrained all filament endpoints within a fixed distance $0.05\cdot D_x$ from the left edge of the domain to be non-moving, then we imposed a rightwards force on all endpoints within a distance $0.05\cdot D_x$ from the right edge of the domain.   To simulate free contraction, we removed all constraints at domain boundaries; to assess buildup and maintenance of contractile stress under isometric conditions, we used periodic boundary conditions in both $x$ and $y$ dimensions.

We measured the local velocity of the network at different positions along the axis of deformation as the mean velocity of all filaments intersecting that position; we measured the internal network stress at each axial position by summing the axial component of the tensions on all filaments intersecting that position, and dividing by network height; finally, we measured network strain rate as the average of all filament velocities divided by their positions.

We assigned biological plausible reference values for all parameters (See Table \ref{table:para}).  For individual analyses, we sampled the ranges of parameter values around these reference values shown in \nameref{S1_Table}.

\begin{table}[h]
\centering
\caption{Simulation parameters with reference values}
\label{table:para}
\begin{tabular}{|c|c|c|c|c|}
\hline
{\bf Parameter}             & {\bf Symbol} & {\bf Reference Value}          \\ \hline
extensional modulus         & $\mu_e$        & $1 nN $                                               \\
compressional modulus             & $\mu_c$     & $ 0.01 nN $                           \\
cross-link drag coefficient & $\xi$      & $unknown $              \\
solvent drag coefficient     & $\zeta$        & $0.0005 \frac{nN s}{\mu m^2} $      \\
filament length             & L            & $5 \mu m$                                          \\
cross-link spacing          & $l_c$        & $0.5 \mu m$                                         \\
active filament force          & $\upsilon$        & $0.1 nN$                                         \\
active cross-link fraction          & $\phi$        & $0.1<0.9$                                         \\
domain size                 & $D_x\times D_y$            & $20\times 50 \mu m$                                 \\ \hline
\end{tabular}
\end{table}

% Results .
\section*{Results}
The goal of this study is to understand how cortical flow is shaped by the simultaneous dependencies of active stress and effective viscosity on filament turnover, crosslink drag and on ``network parameters'' that control  filament density, elasticity and motor activity.   We approach this in three steps: First, we analyze the passive deformation of a cross-linked network in response to an externally applied stress; we identify regimes in which the network response is effectively viscous and characterize the dependence of effective viscosity on network parameters and filament turnover.  Second, we analyze the buildup and dissipation of active stress in cross-linked networks with active motors, as they contract against an external resistance; we identify conditions under which the network can produce sustained stress at steady state, and characterize how steady state stress depends on network parameters and filament turnover. Finally, we confirm that the dependencies of active stress and effective viscosity on network parameters and filament turnover are sufficient to predict the dynamics of networks undergoing steady state flow in response to spatial gradients of motor activity.
% PASSIVE SECTION
\subsection*{Filament turnover allows and tunes effectively viscous steady state flow.}
 
% Example of passive simulation measurements
\paragraph{Networks with passive cross-links and no filament turnover undergo three stages of deformation in response to an extensional force.} 

\begin{figure}[h!]
\centering
\includegraphics[width=\hsize]{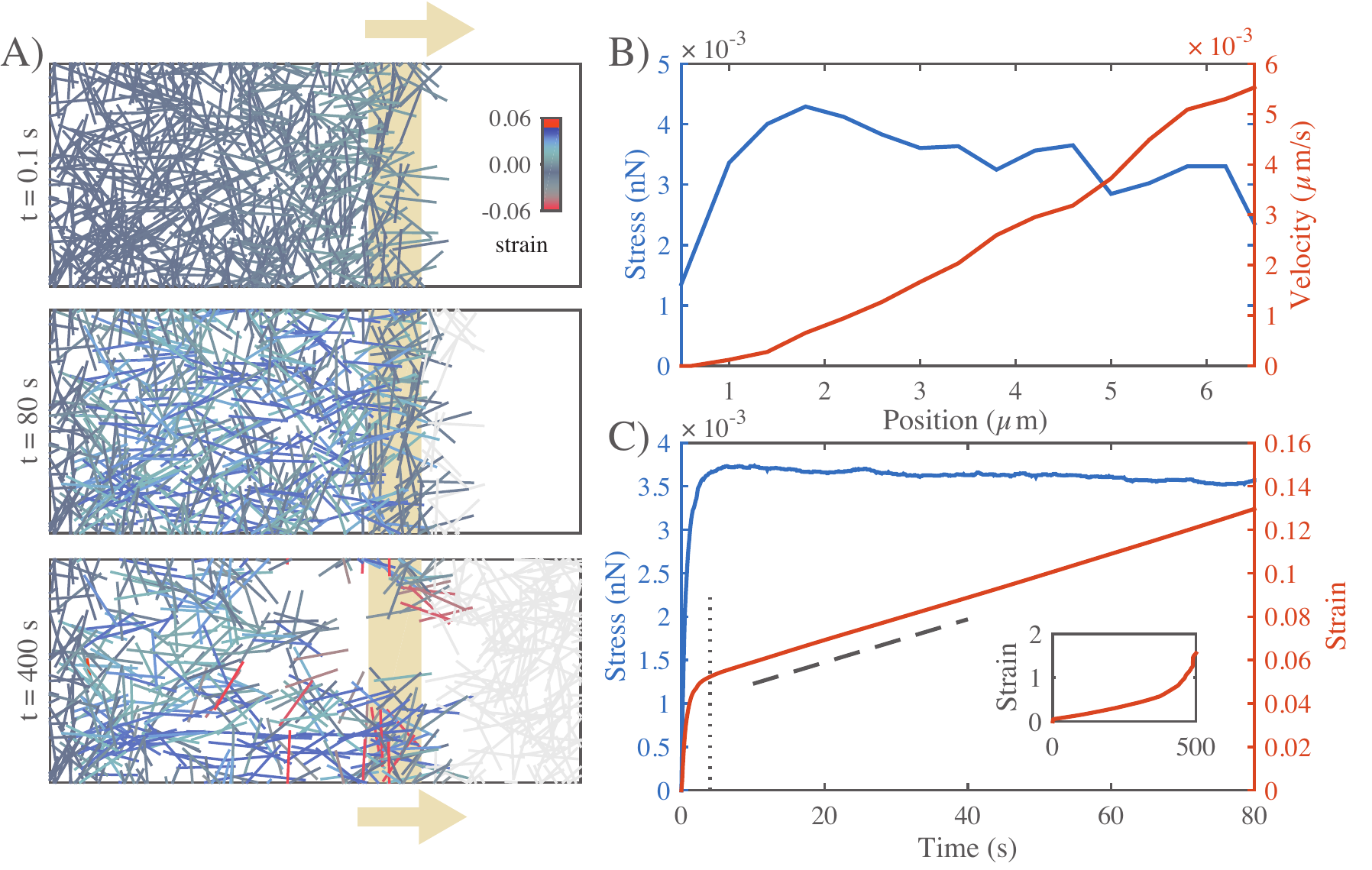}
\caption{\label{fig:passive_ex}  Networks with passive cross-links and no filament turnover undergo three stages of deformation in response to an extensional stress.   \textbf{A)} Three successive time points from a simulation of a $4\times6.6\: \mu m$ network deforming under an applied stress of 0.005 $nN/\mu m$. Stress (tan arrows) is applied to filaments in the region indicated by the tan bar. In this and all subsequent figures, filaments are color-coded with respect to state of strain (blue = tension, red = compression).  Network parameters: $L=1\: \mu m$, $l_c=0.3\: \mu m$, $\xi=100\: nN\cdot s/\mu m$. \textbf{B)} Mean filament stress and velocity profiles for the  network in (a) at t=88s. Note that the stress is nearly constant and the velocity is nearly linear as predicted for a viscous fluid under extension.  \textbf{C)} Plots of the mean stress and strain vs time for the simulation in (a), illustrating the three stages of deformation: (i) A fast initial deformation accompanies rapid buildup of internal network stress; (ii) after a characteristic time $\tau_c$ (indicated by vertical dotted line) the network deforms at a constant rate, i.e. with a constant effective viscosity, $\eta_c$, given by the slope of the dashed line; (iii) at long times, the network undergoes strain thinning and tearing (see inset)}
\end{figure}

To characterize the passive response of a cross-linked filament network without filament recycling, we simulated a simple uniaxial strain experiment in which we pinned the network at one end, imposed an external stress at the opposite end, and then quantified network strain and internal stress as a function of time (Fig. \ref{fig:model_overview}E). The typical response occurred in three qualitatively distinct phases (Fig. \ref{fig:passive_ex}A,C). At short times the response was viscoelastic, with a rapid buildup of internal stress and a rapid $\sim$exponential approach to a fixed strain (\nameref{fig:passive_supp}A), which represents the elastic limit in the absence of cross-link slip predicted by \cite{theo_hlm}. At intermediate times, the local stress and strain rate were approximately constant across the network (Fig. \ref{fig:passive_ex}B), and the response was effectively viscous; internal stress remained ~constant while the network continued to deform slowly and continuously with nearly constant strain rate (shown as dashed line in Fig. \ref{fig:passive_ex}C) as filaments slip past one another against the effective cross-link drag. In this regime, we can quantify effective viscosity, $\eta_c$,  as the ratio of applied stress to the measured strain rate. Finally, as the network strain approached a critical value ($\sim 30\%$ for the simulation in Fig. \ref{fig:passive_ex}), strain thinning lead to decreased network connectivity, local tearing, and rapid acceleration of the network deformation (see inset in Fig. \ref{fig:passive_ex}C).

\begin{figure}[h!]
	\centering
	 \includegraphics[width=\hsize]{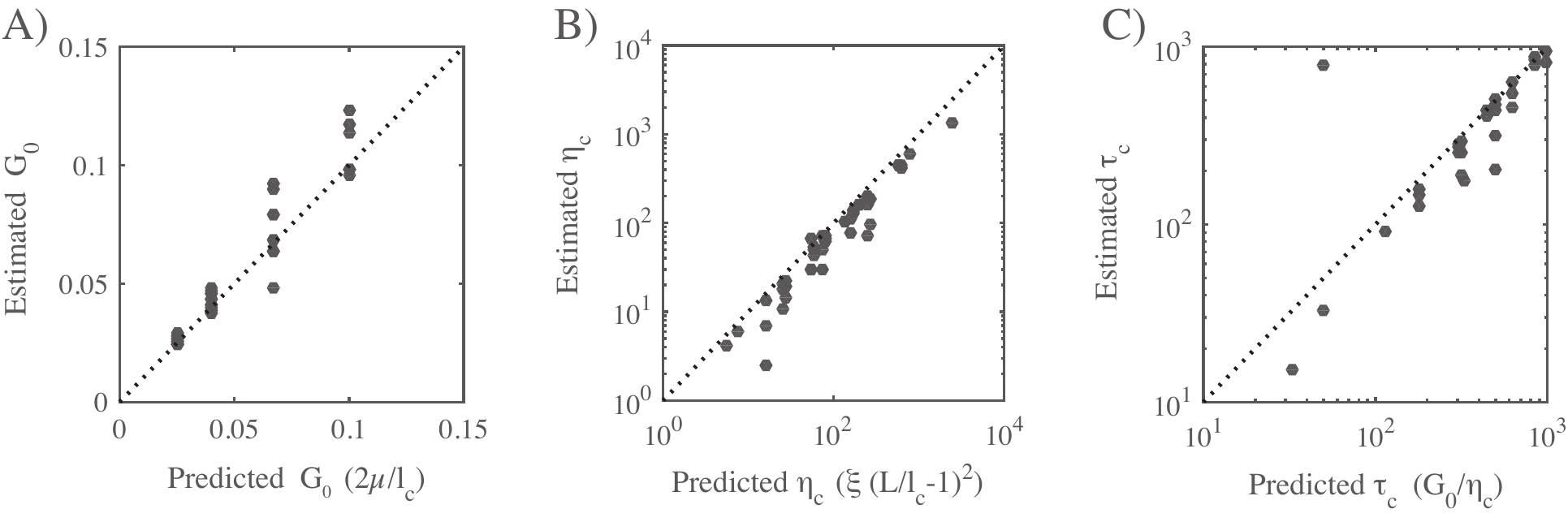}
	\caption{\label{fig:passive_form} Network architecture sets the rate and timescales of deformation.  \textbf{(a-c)} Comparison of predicted and simulated values for: \textbf{A)} the bulk elastic modulus $G_0$,  \textbf{B)} the effective viscosity $\eta_c$ and \textbf{C)} the timescale for transition from viscoelastic to viscous behavior $\tau_c$, given by the ratio of the bulk elastic modulus $G_0$ to effective viscosity, $\eta_c$. Dotted lines indicates the relationships predicted by theory. }
\end{figure}

% Viscosity and timescale parameter dependence
\paragraph{Network architecture sets the rate and timescales of deformation.}  To characterize how effective viscosity and the timescale for transition to effectively viscous behavior depend on network architecture and cross-link dynamics, we simulated a uniaxial stress test, holding the applied stress constant, while varying filament length $L$, density $l_c$,  elastic modulus $\mu_e$ and cross link drag $\xi$ (see \nameref{S1_Table}). We measured the elastic modulus, $G_0$, the effective viscosity, $\eta_c$, and the timescale $\tau_c$ for transition from viscoelastic to effectively viscous behavior, and compared these to theoretical predictions. We observed a transition from viscoelastic to effectively viscous deformation for the entire range of parameter values that we sampled.  Our estimate of $G_0$ from simulation agreed well with the closed form solution  $G_0 \sim \mu/l_c$ predicted by a previous theoretical model \cite{theo_hlm} for networks of semi-flexible filaments with irreversible cross-links (Fig. \ref{fig:passive_form}B). 

A simple theoretical analysis of filament networks with frictional cross link slip, operating in the intermediate viscous regime (see \nameref{S1_Text} A.2), predicted that the effective viscosity $\eta_c$ should be proportional to the cross-link drag coefficient and to the square of the number of cross-links per filament:

\begin{equation}
\eta_c = 4\pi\xi\left ( \frac{L}{l_c}-1\right )^2
\end{equation}

As shown in Fig. \ref{fig:passive_form}B, our simulations agree well with this prediction for a large range of sampled network parameters. Finally, for many linear viscoelastic materials, the ratio of effective viscosity to the elastic modulus $\eta_c/G_0$ sets the timescale for transition from elastic to viscous behavior\cite{mccrum1997principles}. Combining our approximations for $G_0$ and $\eta_c$, we predict a transition time, $\tau_c \approx L^2\xi/l_c\mu$. Measuring the time at which the strain rate became nearly constant (i.e. $\gamma \sim t^n$ with $n>0.8$) yields an estimate of $\tau_c$ that agrees well with this prediction over the entire range of sampled parameters (Fig. \ref{fig:passive_form}C).  Thus the passive response of filament networks with frictional cross link drag is well-described on short (viscoelastic) to intermediate (viscous) timescales by an elastic modulus $G_0$, an effective viscosity $\eta_c$, and a transition timescale $\tau_c$, with well-defined dependencies on network parameters. However, without filament turnover, strain thinning and network tearing limits the extent of viscous deformation to small strains.

% Passive Recycling
\paragraph{Filament turnover allows sustained large-scale viscous flow and defines two distinct flow regimes.}

To characterize how filament turnover shapes the passive network response to an applied force, we introduced a simple form of turnover in which entire filaments disappear at a rate $k_{diss}\rho$, where $\rho$ is the filament density, and new unstrained filaments appear with a fixed rate per unit area, $k_{app}$. In a non-deforming network,  filament density will equilibrate to a steady state value, $\rho_0 = k_{ass}/k_{diss}$, with time constant $\tau_r = 1/k_{diss}$.  However, in networks deforming under extensional stress, the density will be set by a competition between strain thinning and density equilibration via turnover. 

\begin{figure}[h!]
	\centering
		\includegraphics[width=\hsize]{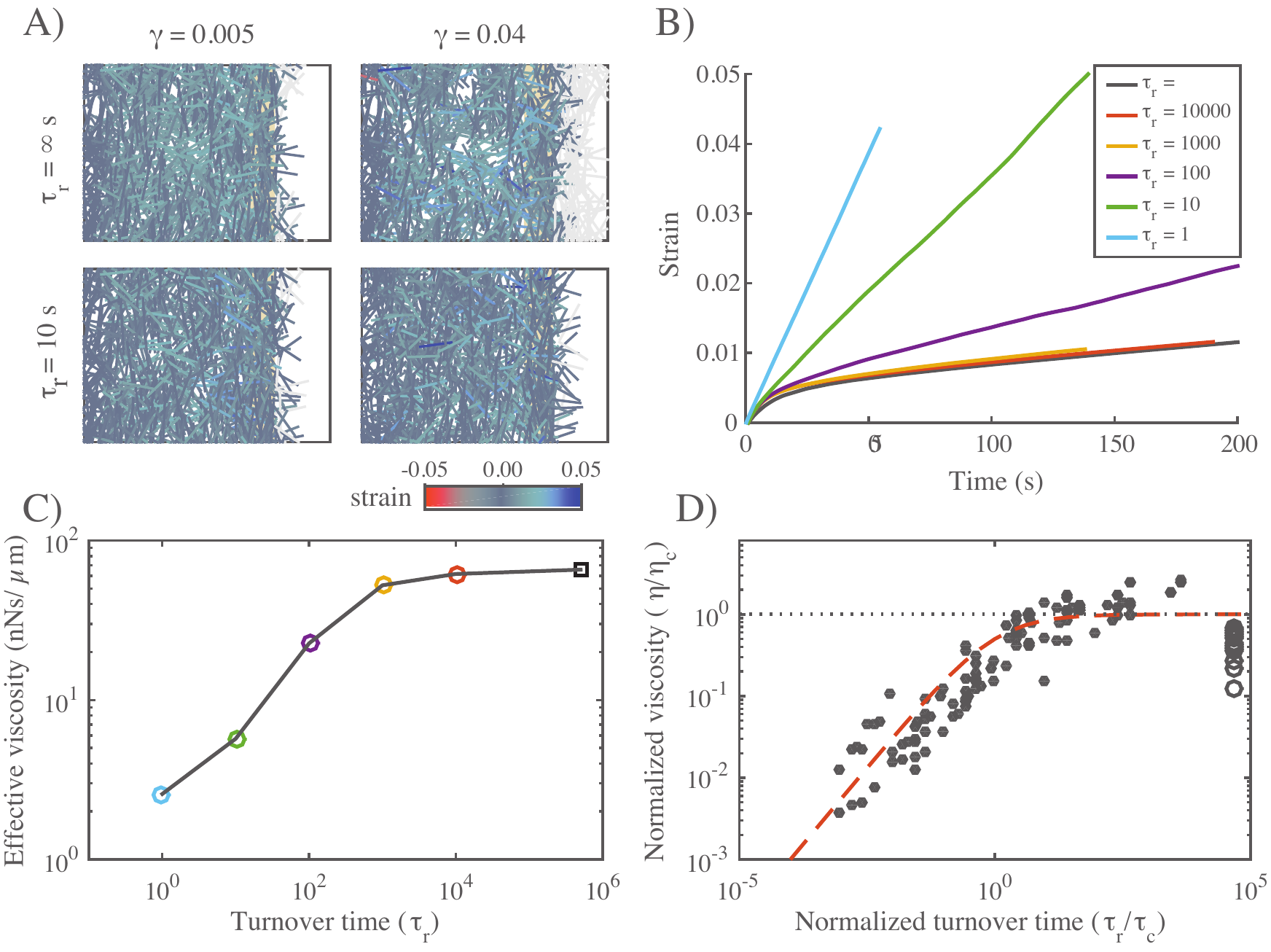}
	\caption{\label{fig:passive_rec}  Filament recycling defines two regimes of effectively viscous flow. \textbf{A)} Comparison of $20 \times 12 \mu m$ networks under 0.001 $nN/\mu m$ extensional stress without (top) and with (bottom) filament turnover.  Both images are taken when the networks had reached a net strain of 0.04.  For clarity, filaments that leave the domain of applied stress are greyed out. \textbf{B)} Plots of strain vs time for identical networks with different rates of filament turnover.  Network parameters: $L=5\: \mu m$, $l_c=0.5\: \mu m$, $\xi=10\: nN\cdot s/\mu m$. \textbf{C)}  Plot of effective viscosity vs turnover time derived from the simulations shown in panel b.  Square dot is the $\tau_r=\infty$ condition.  \textbf{D)} Plot of normalized effective viscosity ($\eta/\eta_c$) vs normalized turnover time ($\tau_r/\tau_c$) for a large range of network parameters and turnover times. For $tau_r \ll \tau_c$, the viscosity of the network becomes dependent on recycling time. Red dashed line indicates the approximation given in equation \ref{eqn:simple_eta} for $m=3/4$.}
\end{figure}

We simulated a uniaxial stress test for different values of $\tau_r$, while holding all other parameters fixed (Fig. \ref{fig:passive_rec}A-C). For large $\tau_r$, as described above, the network undergoes strain thinning and ultimately tears.  Decreasing $\tau_r$ increases the rate at which the network equilibrates towards a steady state density $\rho_0$.  However, it also increases the rate of deformation and thus the rate of strain thinning (Fig. \ref{fig:passive_rec}B).  We found that the former effect dominates, such that below a critical value $\tau_r = \tau_{crit}$, the network can achieve a steady state characterized by a fixed density and a constant strain rate (\nameref{fig:thinning}).  Simple calculations (\nameref{S1_Text} A.3) show that the critical value of $\tau_r$ is approximately:

\begin{equation}
\label{eqn:syst3}
\tau_{crit} = \frac{\xi {\left (\sqrt{\frac{L}{l_c}}-1 \right )}^3}{\sigma}.
\end{equation}

where  $\sigma$ is the applied stress, $L/l_c$ the linear cross link density, and $\xi$ is the effective crosslink drag.

For $\tau_r < \tau_{crit}$, we observed two distinct steady state flow regimes (Fig. \ref{fig:passive_rec}B,C). For intermediate values of $\tau_r$, effective viscosity remains ~constant with decreasing $\tau_r$.  However, below a certain value of  $\tau_r$ ($\approx 10^3$ for the parameters used in Fig. \ref{fig:passive_rec}C),  effective viscosity decreased monotonically with further decreases in $\tau_r$. To understand what sets the timescale for transition between these two regimes,  we measured effective viscosity at steady steady for a wide range of  network parameters ($L, \mu, {l_c}$), crosslink drags ($\xi$) and filament turnover times (Fig. \ref{fig:passive_rec}D). Strikingly, when we plotted the normalized effective viscosity $\eta_r/\eta_c$ vs a normalized recycling rate $\tau_r/\tau_c$ for all parameter values, the data collapsed onto a single curve, with a transition at  $\tau_r \approx \tau_c$  between an intermediate turnover regime in which effective viscosity is independent of $\tau_r$  and an high turnover regime in which effective viscosity falls monotonically with decreasing $~\tau_r/\tau_c$  (Fig. \ref{fig:passive_rec}D). 
 
This biphasic dependence of effective viscosity on filament turnover can be understood intuitively as follows:  As new filaments are born, they become progressively stressed as they stretch and reorient under local influence of surrounding filaments, eventually reaching an elastic limit where their contribution to resisting network deformation is determined by effective crosslink drag.  The time to reach this limit is about the same as the time, $\tau_c$, for an entire network of initially unstrained filaments to reach an elastic limit during the initial viscoelastic response to uniaxial stress, as shown in Fig. 2b.  For $\tau_r < \tau_c$, individual filaments do not have time, on average, to reach the elastic limit before turning over; thus the deformation rate is determined by the elastic resistance of partially strained filaments, which increases with lifetime up to $\tau_r = \tau_c$. For $\tau_r > \tau_c$, the deformation rate is largely determined by cross-link resistance to sliding of maximally strained filaments, and the effective viscosity is insensitive to further increase in  $\tau_r$.

These results complement and extend a previous computational study of irreversibly cross-linked networks of treadmilling filaments deforming under extensional stress\cite{Kim2014526}. Kim et al identified two regimes of effectively viscous deformation: a ``stress-dependent'' regime in which filaments turnover before they become strained to an elastic limit and deformation rate is proportional to both applied stress and turnover rate; and a ``stress-independent'' regime in which filaments reach an elastic limit before turning over and deformation rate depends only on the turnover rate. The fast and intermediate turnover regimes that we observe here correspond to the stress-dependent and independent regimes described by Kim et al, but with a key difference. Without filament turnover, Kim et al's model predicts that a network cannot deform beyond its elastic limit.  In contrast, our model predicts viscous flow at low turnover, governed by an effective viscosity that is set by cross-link density and effective drag. Thus our model provides a self-consistent framework for understanding how crosslink unbinding and filament turnover contribute separately to viscous flow and connects these contributions directly to previous theoretical descriptions of cross-linked networks of semi-flexible filaments. 

In summary, our simulations predict that filament turnover allows networks to undergo viscous deformation indefinitely, without tearing, over a wide range of different effective viscosities and deformation rates. For $\tau_r < \tau_{crit}$, this behavior can be summarized by an equation of the form:

\begin{equation}
\label{eqn:simple_eta}
\eta = \frac{\eta_c}{1+(\tau_c/\tau_r)^m}  
\end{equation}

For $\tau_r \gg \tau_c$, $\eta\approx\eta_c$: effective viscosity depends on crosslink density and effective crosslink drag, independent of changes in recycling rate. For $\tau_r\ll\tau_c$,  effective viscosity is governed by the level of elastic stress on network filaments, and becomes strongly dependent on filament lifetime: $\eta\sim\eta_c(\tau_r/\tau_c)^m$. The origins of the $m = 3/4$ scaling remains unclear (see Discussion).

% ACTIVE SECTION
\subsection*{Filament turnover allows persistent stress buildup in active networks}

\paragraph{In the absence of filament turnover, active networks with free boundaries contract and then stall against passive resistance to network compression.}

\begin{figure}[h!]
	\centering
		\includegraphics[width=\hsize]{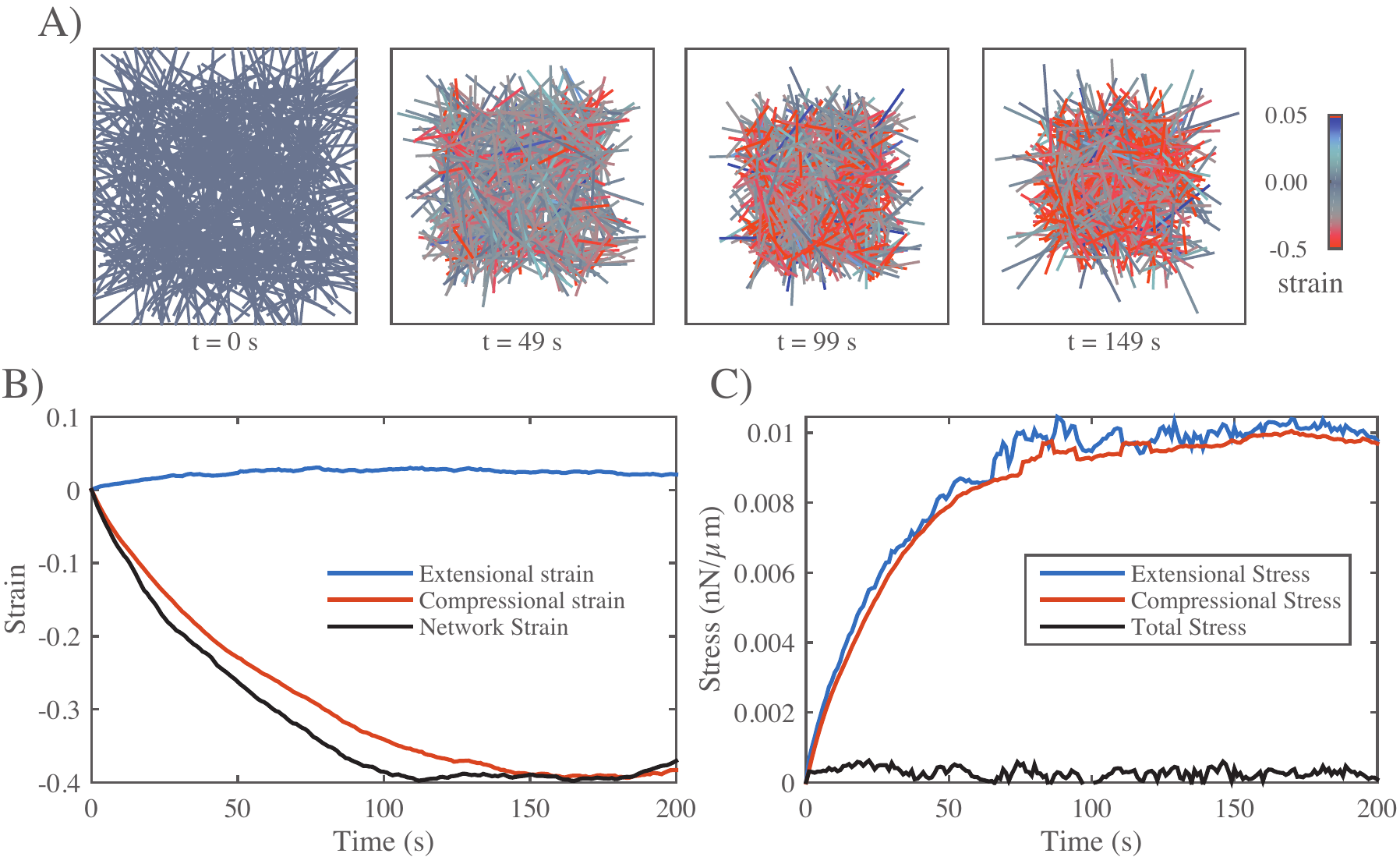}
	\caption{\label{fig:active_con} In the absence of filament turnover, active networks with free boundaries contract and then stall against passive resistance to network compression. \textbf{A)}  Simulation of an active network with free boundaries. Colors represent strain on individual filaments as in previous figures.  Note the buildup of compressive strain as contraction approaches stall between 100 s and 150 s.  Network parameters: $L=5\: \mu m$, $l_c=0.3\: \mu m$, $\xi=1\: nN\cdot s/\mu m$, $\upsilon=0.1\: nN$.  \textbf{B)} Plots showing time evolution of total network strain (black) and the average extensional (blue) or compressive (red) strain on individual filaments.   \textbf{C)} Plots showing time evolution of total (black) extensional (blue) or compressive (red) stress.  Note that extensional and compressive stress remain balanced as compressive resistance builds during network contraction.}
\end{figure}

Previous work \cite{1367-2630-14-3-033037,rheo_2D1,rheo_active} identifies asymmetric filament compliance and spatial heterogeneity in motor activity as minimal requirements for macroscopic contraction of disordered networks. To confirm that our simple implementation of these two requirements (see Models section) is sufficient for macroscopic contraction, we simulated active networks that are unconstrained by external attachments, varying filament length, density, crosslink drag and motor activity.  We observed qualitatively similar results for all choices of parameter values:  Turning on motor activity in an initially unstrained network induced rapid initial contraction, followed by a slower buildup of compressive stress (and strain) on individual filaments, and an $\sim$exponential approach to stall (Fig. \ref{fig:active_con}). The time to stall, $\tau_s$, scaled as $L\xi/\upsilon$ (\nameref{fig:active_supp}A). On even longer timescales, polarity sorting of individual filaments, as previously described \cite{Reymann1310,Murrell15062014,Ndlec:1997aa,Surrey1167} lead to network expansion (see \nameref{active_con_video}).

During the rapid initial contraction, the increase in network strain closely matched the increase in mean compressive strain on individual filaments Fig. \ref{fig:active_con}B, as predicted theoretically \cite{1367-2630-14-3-033037,PhysRevX.4.041002} and observed experimentally\cite{rheo_2D1}. Contraction required asymmetric filament compliance and spatial heterogeneity of motor activity ($\mu_e/\mu_c > 1$, $\phi<1$, \nameref{fig:active_supp}B). Thus our model captures a minimal mechanism for bulk contractility in disordered networks through asymmetric filament compliance and dispersion of motor activity. However, in the absence of turnover, contraction is limited by internal buildup of compressive resistance and the dissipative effects of polarity sorting.

\paragraph{Active networks cannot sustain stress against a fixed boundary in the absence of filament turnover.}

\begin{figure}[h!]
	\centering
	 	\includegraphics[width=\hsize]{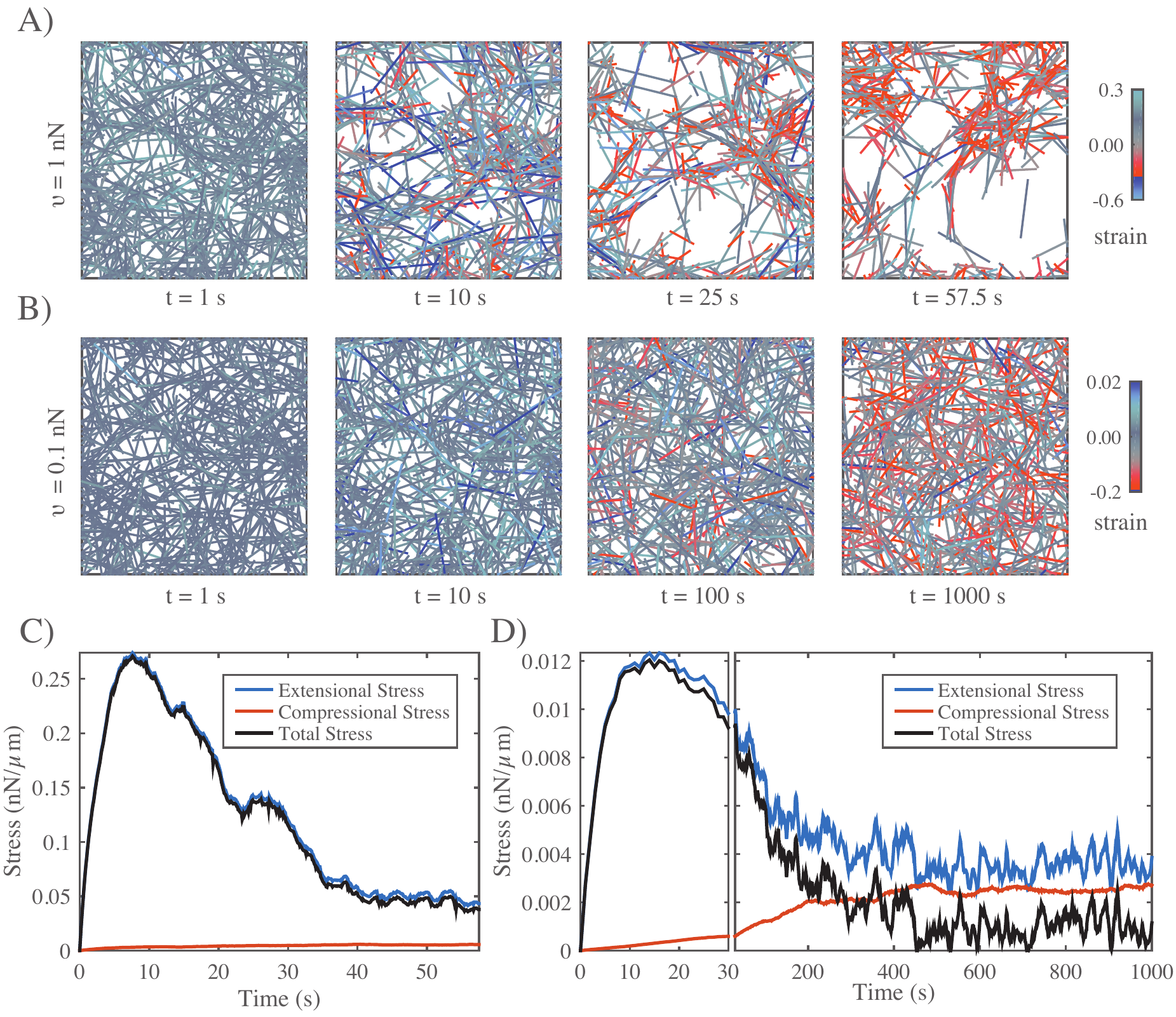}
	\caption{\label{fig:active_str} In the absence of filament turnover, active networks cannot sustain continuous stress against a fixed boundary.  \textbf{A)} Simulation of an active network with fixed boundaries. Rearrangement of network filaments by motor activity leads to rapid loss of network connectivity.  Network parameters: $L=5\: \mu m$, $l_c=0.3\: \mu m$, $\xi=1\: nN\cdot s/\mu m$, $\upsilon=1\: nN$.  \textbf{B)} Simulation of the same network, with the same parameter values, except with ten-fold lower motor activity $\upsilon=0.1\: nN$. In this case, connectivity is preserved, but there is a progressive buildup of compressive strain on individual filaments.  \textbf{C)} Plots of total network stress and the average extensional (blue) and compressive (red) stress on individual filaments for the simulation shown in (a). Rapid buildup of extensional stress allows the network transiently to exert force on its boundary, but this force is rapidly dissipated as network connectivity breaks down.  \textbf{D)} Plots of total network stress and the average extensional (blue) and compressive (red) stress on individual filaments for the simulation shown in (b). Rapid buildup of extensional stress allows the network transiently to exert force on its boundary, but this force is dissipated at longer times as decreasing extensional stress and increasing compressive stress approach balance.  Note the different time scales used for plots and subplots in \textbf{C)} and \textbf{D)} to emphasize the similar timescales for force buildup, but very different timescales for force dissipation.}
\end{figure}

During cortical flow, regions with high motor activity contract against a passive resistance from neighboring regions with lower motor activity.  To understand how the active stresses that drive cortical flow are shaped by motor activity and network remodeling, we analyzed the buildup and maintenance of contractile stress in active networks contracting against a rigid boundary. We simulated active networks contracting from an initially unstressed state against a fixed boundary (Fig. \ref{fig:active_str}A,B), and  monitored the time evolution of mean extensional (blue), compressional (red) and total (black) stress on network filaments (Fig. \ref{fig:active_str}C,D). We focused initially on the scenario in which there is no, or very slow, filament turnover, sampling a range of parameter values controlling filament length and density, motor activity, and crosslink drag. 

We observed a similar behavior in all cases: total stress built rapidly to a peak value $\sigma_m$, and then decayed towards zero (Fig. \ref{fig:active_str}C,D).  The rapid initial increase in total stress was determined largely by the rapid buildup of extensional stress (Fig. \ref{fig:active_str}C,D) on a subset of network filaments (Fig. \ref{fig:active_str}A,B $t=10s$). The subsequent decay involved two different forms of local remodeling: under some conditions, e.g. for higher motor activity (e.g. Fig. \ref{fig:active_str}A,C), the decay was associated with rapid local tearing and fragmentation, leading to global loss of network connectivity as described previously both in simulations\cite{Mak:2016aa} and {\em in vitro}  experiments \cite{Alvarado:2013aa}.  However, for many parameters, (e.g. for higher motor activity  as in Fig. \ref{fig:active_str}B,D), the decay in stress occurred with little or no loss of global connectivity.  Instead, local filament rearrangements changed the balance of extensile vs compressive forces along individual filaments, leading to a slow decrease in the average extensional stress, and a correspondingly slow increase in the compressional stress, on individual filaments (see Fig. \ref{fig:active_str}D).  

Combining dimensional analysis with trial and error, we were able to find empirical scaling relationships describing the dependence of maximum stress $\sigma_m$ and the time to reach maximum stress $\tau_m$ on network parameters and effective crosslink drag  ($\sigma_m \sim \sqrt{\mu_e\upsilon}/l_c$, $\tau_m\sim L\xi/\sqrt{\mu_e\upsilon}$, \nameref{fig:active_supp}C,D). Although these relationships should be taken with a grain of salt, they are reasonably consistent with our simple intuition that the peak stress should increase with motor force ($\upsilon$), extensional modulus ($\mu_e$) and filament density ($1/l_c$), and the time to reach peak stress should increase with crosslink drag ($\xi$) and decrease with motor force ($\upsilon$) and extensional modulus ($\mu_e$).  

\paragraph{Filament turnover allows active networks to exert sustained stress on a fixed boundary.}

\begin{figure}[h!]
	\centering
	 	\includegraphics[width=\hsize]{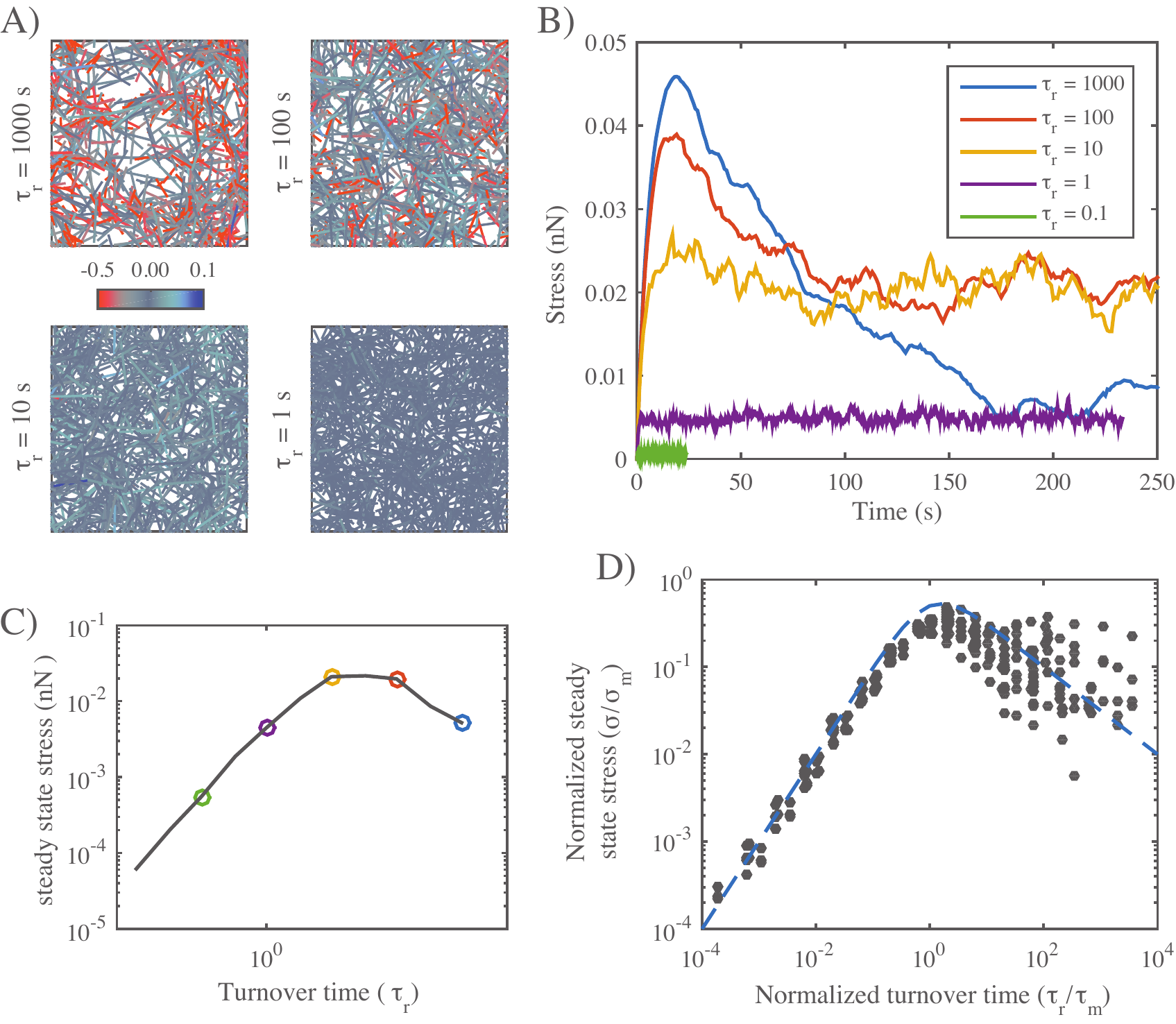}
	\caption{\label{fig:active_rec} Filament turnover allows active networks to exert sustained stress on a fixed boundary. \textbf{A)} Snapshots from simulations of active networks with fixed boundaries and different rates of filament turnover.  All other parameter values are the same as in Fig. \ref{fig:active_str}A. Note the significant buildup of compressive strain and significant remodeling for longer, but not shorter, turnover times. \textbf{B)} Plots of net stress exerted by the network on its boundaries for different recycling times; for long-lived filaments, stress is built rapidly, but then dissipates. Decreasing filament lifetimes reduces stress dissipation by replacing compressed with uncompressed filaments, allowing higher levels of steady state stress; for very short lifetimes, stress is reduced, because individual filaments do not have time to build stress before turning over. \textbf{C)} Plots of $\approx$steady state stress estimated from the simulations in \textbf{B)} vs turnover time.  \textbf{D)} Plot of normalized steady state stress vs normalized recycling time for a wide range of network parameters and turnover times.  Steady state stress is normalized by the predicted maximum stress $\sigma_{m}$ achieved in the absence of filament turnover.  Turnover time is normalized by the predicted time to achieve maximum stress $\tau_{m}$, in the absence of filament turnover.  Predictions for $\sigma_{m}$ and $\tau_{m}$  were obtained from the phenomenological scaling relations shown in (Fig. \ref{fig:active_supp}C,D). Dashed blue line indicates the approximation given in equation \ref{eqn:simple_sigma} for $n=1$.}
\end{figure}

Regardless of the exact scaling dependencies of $\sigma_m$ and $\tau_m$ on network parameters, these results reveal a fundamental limit on the ability of active networks to sustain force against an external resistance in the absence of filament turnover.  To understand how this limit can be overcome by filament turnover, we simulated networks contracting against a fixed boundary from an initially unstressed state, for increasing rates of filament turnover (decreasing $\tau_r$), while holding all other parameter values fixed (Fig. \ref{fig:active_rec}A-C). While the peak stress decreased monotonically with decreasing $\tau_r$, the steady state stress showed a biphasic response, increasing initially with decreasing $\tau_r$, and then falling off as $\tau_r \to 0$.  We observed a biphasic response regardless of how stress decays in the absence of turnover, i.e. whether decay involves loss of network connectivity, or local remodeling without loss of connectivity, or both (\nameref{fig:active_tear} and not shown). Significantly, when we plot normalized steady state stress ($\sigma/\sigma_m$) vs normalized recycling time ($\tau_r$/$\tau_m$) for a wide range of network parameters, the data collapse onto a single biphasic response curve, with a peak near $\tau_r/\tau_m = 1$ (Fig. \ref{fig:active_rec}D). In particular, for $\tau_r < \tau_m$, the scaled data collapsed tightly onto a single curve representing a linear increase in steady state stress with increasing $\tau_r$. For $\tau_r > \tau_m$, the scaling was less consistent, although the trend towards a monotonic decrease with increasing $\tau_r$ was clear. These results reveal that filament turnover can ``rescue" the dissipation of active stress during isometric contraction due to network remodeling, and they show that, for a given choice of network parameters, there is an optimal choice of filament lifetime that maximizes steady state stress.

We can understand the biphasic dependence of steady state stress on filament lifetime using the same reasoning applied to the case of passive flow:   During steady state contraction, the average filament should build and dissipate active stress on approximately the same schedule as an entire network contracting from an initially unstressed state (Fig. \ref{fig:active_rec}B). Therefore for $\tau_r < \tau_m$, increasing lifetime should increase the mean stress contributed by each filament. For $\tau_r > \tau_m$, further increases in lifetime should begin to reduce the mean stress contribution. Directly comparing the time-dependent buildup and dissipation of stress in the absence of turnover, with the dependence of steady state stress on $\tau_r$, supports this interpretation (\nameref{fig:recycle_supp})

As for the passive response (i.e. Equation \ref{eqn:simple_eta}), we can describe this biphasic dependence phenomenologically with an equation of the form:

\begin{equation}
\label{eqn:simple_sigma}
\sigma_{ss} = \frac{\sigma_m}{(\tau_r/\tau_m)^n+\tau_m/\tau_r}  
\end{equation}

where the origins of the exponent $n$ remain unclear.

% COMBINED SECTION
\subsection*{Filament turnover tunes the balance between active stress buildup and viscous stress relaxation to generate flows}

\begin{figure}[h!]
	\centering
	 	\includegraphics[width=\hsize]{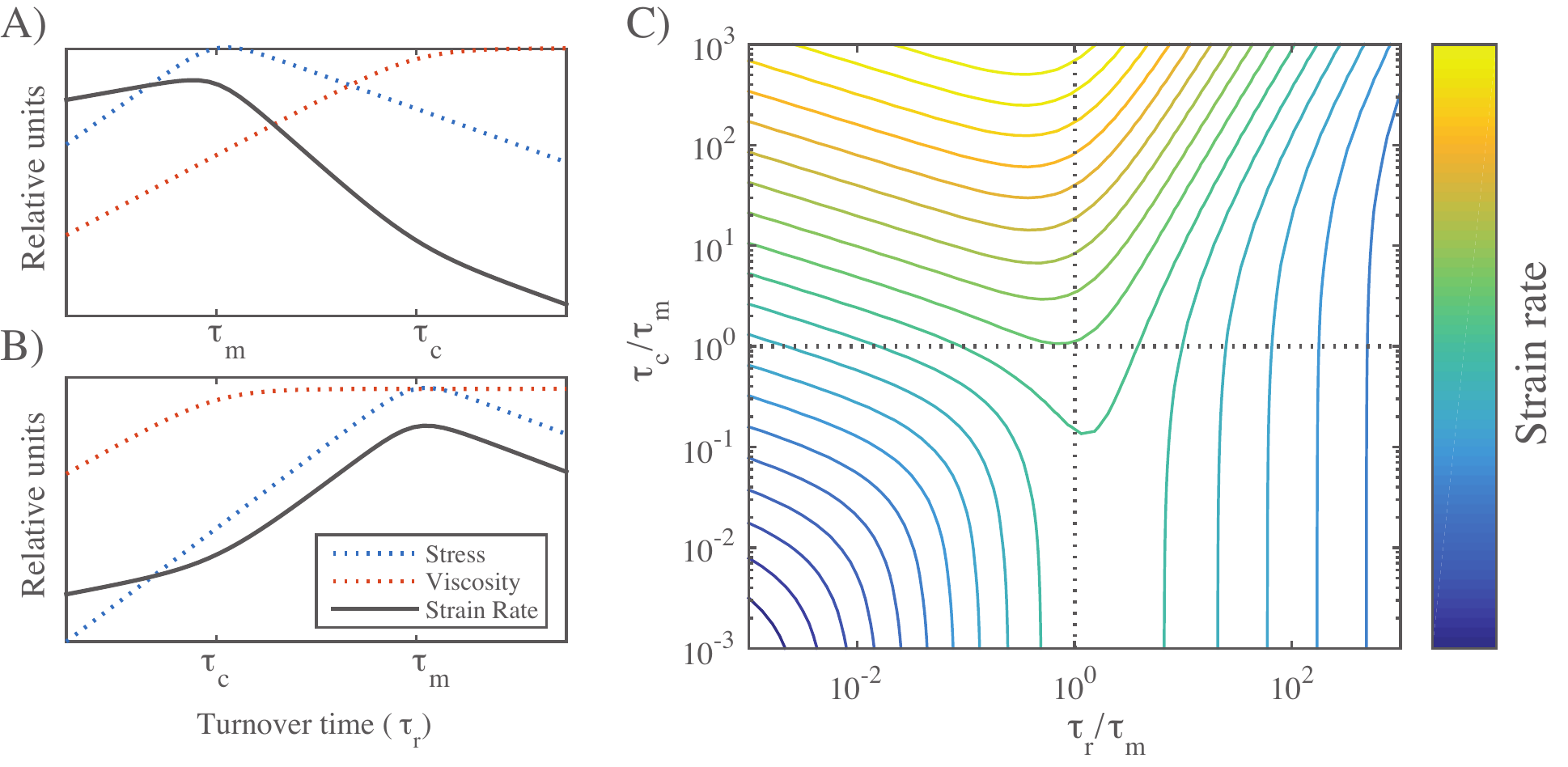}
	\caption{\label{fig:flow_theo}  Filament recycling tunes the magnitudes of both effective viscosity and steady state stress. \textbf{A)}  Dependence of steady state stress, effective viscosity, and resulting strain rate on recycling time $\tau_r$ under the condition $\tau_{m}<\tau_c$. \textbf{B)} Same as a) but for $\tau_c<\tau_{m}$.  \textbf{C)} State space of flow rate dependence relative to the two relaxation timescales, $\tau_r$ and $\tau_c$ normalized by the stress buildup timescale, $\tau_{m}$.  }
\end{figure}

Thus far, we have considered independently how network remodeling controls the passive response to an external stress, or the steady state stress produced by active contraction against an external resistance. We now consider how these two forms of dependence will combine to shape steady state flow produced by spatial gradients of motor activity. We consider a simple scenario in which a network is pinned on either side and motor activity is continuously patterned such that the right half network has uniformly high levels of motor activity (controlled by $\upsilon$, with $\psi = 0.5$), while the left half network has none. Under these conditions, the right half network will contract continuously against a passive resistance from the left half network.  Because of asymmetric filament compliance, the internal resistance of the right half network to active compression should be negligible compared to the external resistance of the left half network.  Thus the steady state flow will be described by:

\begin{equation}
\label{eqn:everybody_knows}
\dot{\gamma} = \frac{\sigma_{ss}}{\eta}  
\end{equation}

where $\sigma_{ss}$ is the active stress generated by the right half-network (less the internal resistance to filament compression), $\eta$ is the effective viscosity of the left half network and strain rate $\dot{\gamma}$ is measured in the left half-network.  Note that strain rate can be related to the steady state flow velocity $v$ at the boundary between right and left halves through $ v = \dot{\gamma}Dx$. Therefore, we can understand the dependence of flow speed on filament turnover and other parameters using the approximate relationships summarized by equations \ref{eqn:simple_eta} and \ref{eqn:simple_sigma} for $\eta$ and $\sigma_{ss}$.  As shown in Fig. \ref{fig:flow_theo}, there are two qualitatively distinct possibilities for the dependence of strain rate on $\tau_r$, depending on the relative magnitudes of $\tau_m$ and $\tau_c$.  In both cases, for fast enough turnover ($\tau_r < \min \left (\tau_m, \tau_c \right )$), we expect weak dependence of strain rate on $\tau_r$ ($ \dot{\gamma}\sim \tau_r^{1/4}$).  For all parameter values that we sampled in this study (which were chosen to lie in a physiological range), $\tau_m > \tau_c$. Therefore we predict the dependence of steady state strain rate on $\tau_r$ shown in Fig. \ref{fig:flow_theo}A.

\begin{figure}[h!]
	\centering
	 	\includegraphics[width=\hsize]{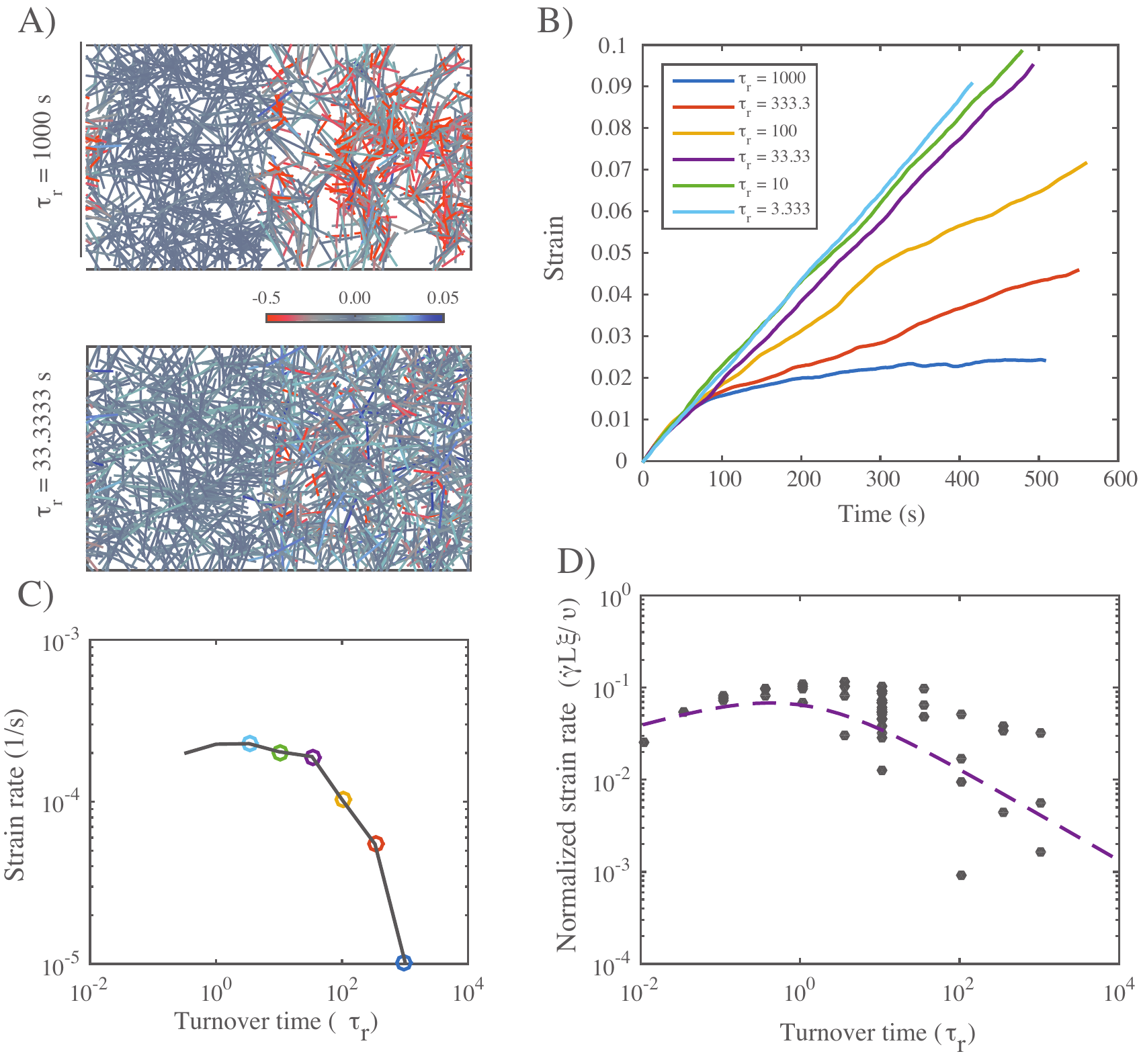}
	\caption{\label{fig:flow_ex}  Filament recycling allows sustained flows in networks with non-isotropic activity. \textbf{A)} Example simulations of non-isotropic networks with long ($\tau_r=1000$) and short ($\tau_r=33$) recycling timescales. In these networks the left half of the network is passive while the right half is active.  Network parameters are same as in Fig.s \ref{fig:active_str} and \ref{fig:active_rec}. Importantly, in all simulations $\tau_{m}<\tau_c$. \textbf{B)} Graph of strain for identical networks with varying recycling timescales.  With long recycling times, the network stalls; reducing the recycling timescale allows the network to persist in its deformation.  However, for the shortest recycling timescales, the steady state strain begins to slowly return to 0 net motion.  Measurements are based on the passive side of the network. \textbf{C)} Steady state strain rates for the networks in (b).     \textbf{D)} Graph of network's long-term strain rate as a function of recycling timescale. Dashed line is form of dependence predicted by the  theoretical arguments shown in Fig. \ref{fig:flow_theo}.}
\end{figure}

To confirm this prediction, we simulated the simple scenario described above for a range of values of $\tau_r$, with all other parameter values initially fixed. As expected, we observed a sharp dependence of steady state flow speeds on filament recycling rate (Fig. \ref{fig:flow_ex}B,C). For very long recycling times, ($\tau_r=1000 s$, dark blue line), there was a rapid initial deformation (contraction of the active domain and dilation of the passive domain), followed by a slow approach to a steady state flow characterized by slow contraction of the right half-domain and a matching dilation of the left half-domain (see \nameref{fig:combo_prof}).  However, with decreasing values of $\tau_r$, steady state flow speeds increased steadily, before reaching an approximate plateau on which flow speeds varied by less than 15 \% over more than two decades of variation in $\tau_r$ (Fig. \ref{fig:flow_ex}C).  

We repeated these simulations for a wider range of parameter values, and saw similar dependence of $\dot{\gamma}$ on $\tau_r$ in all cases.  Using equation \ref{eqn:simple_eta} with $\tau_r < \tau_c$ and equation \ref{eqn:simple_sigma} with $\tau_r < \tau_m$, and the theoretical or empirical scaling relationships found above for $\eta_c$, $\tau_c$, $\sigma_m$ and $\tau_m$, we predict a simple scaling relationship for $\dot{\gamma}$ (for small $\tau_r$):

\begin{equation}
\label{eqn:flow_scaling_eq}
\dot{\gamma} = \frac{\upsilon}{\xi L}  \left ( \tau_r \right ) ^{1/4}
\end{equation}

Indeed, when we plot the steady state measurements of $\dot{\gamma}$, normalized by $\upsilon/\xi L$,  for all parameter values, the data collapse onto a single curve for small $\tau_r$.  Thus. our simulations identify a flow regime, characterized by sufficiently fast filament turnover, in which the steady state flow speed is buffered against variation in turnover, and has a relatively simple dependence on other network parameters.

%Conclusion
\section*{Discussion}

\paragraph{} Cortical flows are shaped by the dynamic interplay of force production and dissipation within cross-linked actomyosin networks. Here we combined computational models with simple theoretical analyses to explore how this interplay depends on filament turnover, crosslink dynamics and network architecture. Our results reveal an essential requirement for filament turnover during cortical flow, both to sustain active stress and to continuously relax elastic resistance without catastrophic loss of network connectivity. Moreover, we find that biphasic dependencies of active stress and passive relaxation on filament lifetime define multiple regimes for steady state flow with distinct dependencies on network parameters and filament turnover.

\paragraph{} We identify two regimes of passive response to external stress:  a low turnover regime in which filaments strain to an elastic limit before turning over, and effective viscosity depends on crosslink density and effective crosslink friction, and a high turnover regime in which filaments turn over before reaching an elastic limit and effective viscosity is proportional to elastic resistance and $\sim$proportional to filament lifetime. Thus our model captures the qualitatively distinct contributions of transient crosslinks and filament turnover within a single self-consistent framework.  We note that the weakly sub-linear dependence of effective viscosity on filament lifetime that we observe here may simply reflect a failure to capture very local modes of filament deformation, since a previous study \cite{Kim2014526} in which filaments were represented as connected chains of smaller segments predicted linear dependence of effective viscosity on filament lifetime.

\paragraph{} Our simulations active networks confirm the theoretical prediction \cite{1367-2630-14-3-033037,rheo_2D1,rheo_active} that spatial heterogeneity of motor activity and asymmetric filament compliance are sufficient to support macroscopic contraction of unconstrained networks. However, under isometric conditions, and without filament turnover, our simulations predict that active stress cannot be sustained. On short timescales, motor forces drive local buildup of extensional stress, but on longer timescales, local motor-driven filament rearrangements and thus local changes in connectivity, invariably lead to a decay in active stress.  Under some conditions, contractile forces drive networks towards a critically connected state, leading to tearing and fragmentation, as previously described \cite{Alvarado:2013aa, Mak:2016aa}. However, we find that stress decay can also occur without any global loss of connectivity, through a gradual decrease in extensile force and a gradual increase in compressive force along individual filaments.   When filaments can slide relative to one another, the motor forces that produce active stress will inevitably lead to local changes in connectivity that decrease active stress.  These results suggest that for contractile networks to maintain isometric tension on long timescales, they must either form stable crosslinks to prevent filament rearrangements, or they must continuously recycle network filaments (or active motors) to renew the local potential for production of active stress.

\paragraph{} Indeed, our simulations predict that filament turnover is sufficient for maintenance of active stress and they predict a biphasic dependence of steady state stress on filament turnover: For short-lived filaments ($\tau_r < \tau_m$), steady state stress increases linearly with filament lifetime because filaments have more time to build towards peak extensional stress before turning over.  For longer loved filaments ($\tau_r > \tau_m$), steady state stress decreases monotonically with filament lifetime because local rearrangements decrease the mean contributions of longer lived filaments. These findings imply that for cortical networks that sustain contractile stress under approximately isometric conditions, tuning filament turnover can control the level of active stress, and there will be an optimal turnover rate that maximizes the stress, all other things equal.  This may be important, for example in early development, where contractile forces produced by cortical actomyosin networks play key roles in maintaining, or controlling slow changes in cell shape and tissue geometry \cite{Salbreux2012536,Turlier2014114,Gorfinkiel2011531}.

\paragraph{} For cortical networks that undergo steady state flows driven by spatial gradients of motor activity, our simulations predict that the biphasic dependencies of steady state stress and effective viscosity on filament lifetime define multiple regimes of steady state flow, characterized by different dependencies on filament turnover (and other network parameters).  In particular, the $\sim$linear dependencies of steady state stress and effective viscosity on filament lifetime for short-lived filaments define a fast turnover regime in which steady state flow speeds are buffered against variations in filament lifetime, and are predicted to depend in a simple way on motor activity and crosslink resistance.  Measurements of F-actin turnover times in cells that undergo cortical flow \cite{Theriot1991,Murthy2016,Watanabe1083,Guha2016,Fritzsche15032013,Robin:2014aa} suggests that they may indeed operate in this fast turnover regime, and recent studies in {\em C. elegans} embryos suggests that cortical flow speeds are surprisingly insensitive to depletion of factors (ADF/Cofilin) that govern filament turnover \cite{cellmech_flows}, consistent with our model's predictions. Stronger tests of our model's predictions will require more systematic analyses of how flow speeds vary with filament and crosslink densities, motor activities, and filament lifetimes. 

\section*{Supporting Information}

% Include only the SI item label in the subsection heading. Use the \nameref{label} command to cite SI items in the text.

\paragraph*{S1 Appendix.}
\label{S1_Text}
{\bf Code Reference and Supplementary Methods}  \textbf{A.1)} Reference to simulation and analysis code. \textbf{A.2)} Derivation of effective viscosity. \textbf{A.3)} Derivation of critical turnover timescale for steady state flow

\paragraph*{S1 Table.}
\label{S1_Table}
{\bf Parameter values.}  List of parameter values used for each set of simulations.

\paragraph*{S1 Fig.}
\label{fig:passive_supp}
{\bf  Fast viscoelastic response to extensional stress.}  Plots of normalized strain vs time during the elastic phase of deformation in passive networks under extensional stress.  Measured strain is normalized by the equilibrium strain predicted for a network of elastic filaments without crosslink slip $\gamma_{eq} = \sigma/G_0 = \sigma/(2\mu/l_c)$.

\paragraph*{S2 Fig.}
\label{fig:thinning}
{\bf  Filament turnover rescues strain thinning.} \textbf{A)} Plots of strain vs time for different turnover times (see inset in (b)). Note the increase in strain rates with decreasing turnover time. \textbf{B)} Plots of filament density vs strain for different turnover times $\tau_r$.  For intermediate $\tau_r$, simulations predict progressive strain thinning, but at a lower rate than in the complete absence of recycling. For higher $\tau_r$, densities approach steady state values at longer times.  

\paragraph*{S3 Fig.}
\label{fig:active_supp}
{\bf  Mechanical properties of active networks.}  \textbf{A)}  Time for freely contracting networks to reach maximum strain, $\tau_s$, scales with $L\xi/\upsilon$.  \textbf{B)} Free contraction requires asymmetric filament compliance, and total network strain increases with the applied myosin force $\upsilon$. Note that the maximum contraction approaches an asymptotic limit as the stiffness asymmetry approaches a ratio of $\sim 100$.   \textbf{C)}  Maximum stress achieved during isometric contraction, $\sigma_m$, scales approximately with $\sqrt{\mu_e\upsilon}/l_c$.  \textbf{D)} Time to reach max stress during isometric contraction scales approximately with $L\xi/\sqrt{\mu_e\upsilon}$. Scalings for $\tau_s$, $\sigma_m$ and $\tau_m$ were determined empirically by trial and error, guided by dimensional analysis.  

\paragraph*{S4 Fig.}
\label{fig:active_tear}
{\bf  Filament turnover prevents tearing of active networks.}  Plots of normalized strain vs time during the elastic phase of deformation in passive networks under extensional stress.  Measured strain is normalized by the equilibrium strain predicted for a network of elastic filaments without crosslink slip $\gamma_{eq} = \sigma/G_0 = \sigma/(2\mu/l_c)$.

\paragraph*{S5 Fig.}
\label{fig:recycle_supp}
{\bf Bimodal dependence on turnover time matches bimodal buildup and dissipation of stress in the absence of turnover.}   \textbf{A)}  Bimodal buildup of stress in a network with very slow turnover ($\tau_r = 1000s$).  \textbf{B)}  Steady state stress for networks with same parameters as in (a), but  for a range of filament turnover times.   

\paragraph*{S6 Fig.}
\label{fig:combo_prof}
{\bf  Dynamics of steady state flow.}  Plots of stress and strain vs position for networks in which motor activity is limited to the right-half domain and filament turnover time is either  \textbf{A)} $\tau_r = 10000$ or  \textbf{B)} $\tau_r = 10 s$.  Blue indicates velocity while orange represents total stress, measured as described in the main text.

\paragraph*{S1 Video.}
\label{passive_ex_video}
{\bf Extensional strain in passive networks.}  Movie of simulation setup shown in Fig. \ref{fig:passive_ex}.  Colors are the same as in figure.

\paragraph*{S2 Video.}
\label{active_con_video}
{\bf Active networks contracting with free boundaries.}  Movie of simulation setup shown in Fig. \ref{fig:active_con}.  Colors are the same as in figure.

\section*{Acknowledgments}
We would like to thank Shiladitya Banerjee for stimulating discussions.

\nolinenumbers

%\section*{References}
% Compile your BiBTeX database using our plos2015.bst
% style file and paste the contents of your .bbl file
% here.
% 
\bibliographystyle{plos2015}
\bibliography{slippage,active}

\begin{thebibliography}{10}

\bibitem{cellmech_flows3}
Bray D, White J.
\newblock Cortical flow in animal cells.
\newblock Science. 1988;239(4842):883--888.
\newblock Available from:
  \url{http://www.sciencemag.org/content/239/4842/883.abstract}.

\bibitem{cellmech_flows2}
Hird SN, White JG.
\newblock Cortical and cytoplasmic flow polarity in early embryonic cells of
  Caenorhabditis elegans.
\newblock The Journal of Cell Biology. 1993;121(6):1343--1355.
\newblock Available from:
  \url{http://jcb.rupress.org/content/121/6/1343.abstract}.

\bibitem{Benink:2000aa}
Benink HA, Mandato CA, Bement WM.
\newblock Analysis of Cortical Flow Models In Vivo.
\newblock Molecular Biology of the Cell. 2000 08;11(8):2553--2563.
\newblock Available from:
  \url{http://www.ncbi.nlm.nih.gov/pmc/articles/PMC14939/}.

\bibitem{Wilson:2010aa}
Wilson CA, Tsuchida MA, Allen GM, Barnhart EL, Applegate KT, Yam PT, et~al.
\newblock Myosin II contributes to cell-scale actin network treadmilling
  through network disassembly.
\newblock Nature. 2010 05;465(7296):373--377.
\newblock Available from: \url{http://dx.doi.org/10.1038/nature08994}.

\bibitem{Rauzi2010}
Rauzi M, Lenne PF, Lecuit T.
\newblock Planar polarized actomyosin contractile flows control epithelial
  junction remodelling.
\newblock Nature. 2010 Dec;468(7327):1110--1114.
\newblock Available from: \url{http://dx.doi.org/10.1038/nature09566}.

\bibitem{Munro2004413}
Munro E, Nance J, Priess JR.
\newblock Cortical Flows Powered by Asymmetrical Contraction Transport \{PAR\}
  Proteins to Establish and Maintain Anterior-Posterior Polarity in the Early
  C. elegans Embryo.
\newblock Developmental Cell. 2004;7(3):413 -- 424.
\newblock Available from:
  \url{http://www.sciencedirect.com/science/article/pii/S153458070400276X}.

\bibitem{Salbreux2012536}
Salbreux G, Charras G, Paluch E.
\newblock Actin cortex mechanics and cellular morphogenesis.
\newblock Trends in Cell Biology. 2012;22(10):536 -- 545.
\newblock Available from:
  \url{http://www.sciencedirect.com/science/article/pii/S0962892412001110}.

\bibitem{Murrell:2015aa}
Murrell M, Oakes PW, Lenz M, Gardel ML.
\newblock Forcing cells into shape: the mechanics of actomyosin contractility.
\newblock Nat Rev Mol Cell Biol. 2015 08;16(8):486--498.
\newblock Available from: \url{http://dx.doi.org/10.1038/nrm4012}.

\bibitem{Bendix20083126}
Bendix PM, Koenderink GH, Cuvelier D, Dogic Z, Koeleman BN, Brieher WM, et~al.
\newblock A Quantitative Analysis of Contractility in Active Cytoskeletal
  Protein Networks.
\newblock Biophysical Journal. 2008;94(8):3126 -- 3136.
\newblock Available from:
  \url{http://www.sciencedirect.com/science/article/pii/S0006349508704697}.

\bibitem{Janson1005}
Janson LW, Kolega J, Taylor DL.
\newblock Modulation of contraction by gelation/solation in a reconstituted
  motile model.
\newblock The Journal of Cell Biology. 1991;114(5):1005--1015.
\newblock Available from: \url{http://jcb.rupress.org/content/114/5/1005}.

\bibitem{cellmech_flows}
Mayer M, Depken M, Bois JS, Julicher F, Grill SW.
\newblock Anisotropies in cortical tension reveal the physical basis of
  polarizing cortical flows.
\newblock Nature. 2010 09;467(7315):617--621.
\newblock Available from: \url{http://dx.doi.org/10.1038/nature09376}.

\bibitem{PhysRevLett.106.028103}
Bois JS, J\"ulicher F, Grill SW.
\newblock Pattern Formation in Active Fluids.
\newblock Phys Rev Lett. 2011 Jan;106:028103.
\newblock Available from:
  \url{http://link.aps.org/doi/10.1103/PhysRevLett.106.028103}.

\bibitem{De-La-Cruz:2015aa}
De~La~Cruz EM, Gardel ML.
\newblock Actin Mechanics and Fragmentation.
\newblock Journal of Biological Chemistry. 2015 07;290(28):17137--17144.
\newblock Available from: \url{http://www.jbc.org/content/290/28/17137.abstract
  N2 - Cell physiological processes require the regulation and coordination of
  both mechanical and dynamical properties of the actin cytoskeleton. Here we
  review recent advances in understanding the mechanical properties and
  stability of actin filaments and how these properties are manifested at
  larger (network) length scales. We discuss how forces can influence local
  biochemical interactions, resulting in the formation of mechanically
  sensitive dynamic steady states. Understanding the regulation of such
  force-activated chemistries and dynamic steady states reflects an important
  challenge for future work that will provide valuable insights as to how the
  actin cytoskeleton engenders mechanoresponsiveness of living cells.}

\bibitem{Turlier2014114}
Turlier H, Audoly B, Prost J, Joanny JF.
\newblock Furrow Constriction in Animal Cell Cytokinesis.
\newblock Biophysical Journal. 2014;106(1):114 -- 123.
\newblock Available from:
  \url{http://www.sciencedirect.com/science/article/pii/S0006349513012447}.

\bibitem{PhysRevLett.103.058102}
Salbreux G, Prost J, Joanny JF.
\newblock Hydrodynamics of Cellular Cortical Flows and the Formation of
  Contractile Rings.
\newblock Phys Rev Lett. 2009 Jul;103:058102.
\newblock Available from:
  \url{http://link.aps.org/doi/10.1103/PhysRevLett.103.058102}.

\bibitem{Keren:2009aa}
Keren K, Yam PT, Kinkhabwala A, Mogilner A, Theriot JA.
\newblock Intracellular fluid flow in rapidly moving cells.
\newblock Nat Cell Biol. 2009 10;11(10):1219--1224.
\newblock Available from: \url{http://dx.doi.org/10.1038/ncb1965}.

\bibitem{RevModPhys.85.1143}
Marchetti MC, Joanny JF, Ramaswamy S, Liverpool TB, Prost J, Rao M, et~al.
\newblock Hydrodynamics of soft active matter.
\newblock Rev Mod Phys. 2013 Jul;85:1143--1189.
\newblock Available from:
  \url{http://link.aps.org/doi/10.1103/RevModPhys.85.1143}.

\bibitem{Behrndt257}
Behrndt M, Salbreux G, Campinho P, Hauschild R, Oswald F, Roensch J, et~al.
\newblock Forces Driving Epithelial Spreading in Zebrafish Gastrulation.
\newblock Science. 2012;338(6104):257--260.
\newblock Available from:
  \url{http://science.sciencemag.org/content/338/6104/257}.

\bibitem{rheo_fluid}
Hochmuth RM.
\newblock Micropipette aspiration of living cells.
\newblock Journal of Biomechanics. 2000;33(1):15 -- 22.
\newblock Available from:
  \url{http://www.sciencedirect.com/science/article/pii/S002192909900175X}.

\bibitem{rheo_fluid2}
Evans E, Yeung A.
\newblock Apparent viscosity and cortical tension of blood granulocytes
  determined by micropipet aspiration.
\newblock Biophysical Journal. 1989 07;56(1):151--160.
\newblock Available from:
  \url{http://www.ncbi.nlm.nih.gov/pmc/articles/PMC1280460/}.

\bibitem{cell_rheo_exp}
Bausch AR, Ziemann F, Boulbitch AA, Jacobson K, Sackmann E.
\newblock Local Measurements of Viscoelastic Parameters of Adherent Cell
  Surfaces by Magnetic Bead Microrheometry.
\newblock Biophysical Journal. 1998;75(4):2038 -- 2049.
\newblock Available from:
  \url{http://www.sciencedirect.com/science/article/pii/S0006349598776465}.

\bibitem{De-La-Cruz:2009aa}
De~La~Cruz EM.
\newblock How cofilin severs an actin filament.
\newblock Biophysical reviews. 2009 05;1(2):51--59.
\newblock Available from:
  \url{http://www.ncbi.nlm.nih.gov/pmc/articles/PMC2917815/}.

\bibitem{theo_crosslinkslip1}
Broedersz CP, Depken M, Yao NY, Pollak MR, Weitz DA, MacKintosh FC.
\newblock Cross-Link-Governed Dynamics of Biopolymer Networks.
\newblock Phys Rev Lett. 2010 Nov;105:238101.
\newblock Available from:
  \url{http://link.aps.org/doi/10.1103/PhysRevLett.105.238101}.

\bibitem{theo_crosslinkslip2}
M\"uller KW, Bruinsma RF, Lieleg O, Bausch AR, Wall WA, Levine AJ.
\newblock Rheology of Semiflexible Bundle Networks with Transient Linkers.
\newblock Phys Rev Lett. 2014 Jun;112:238102.
\newblock Available from:
  \url{http://link.aps.org/doi/10.1103/PhysRevLett.112.238102}.

\bibitem{model_taeyoon}
Kim T, Hwang W, Kamm RD.
\newblock Dynamic Role of Cross-Linking Proteins in Actin Rheology.
\newblock Biophysical Journal. 2011 10;101(7):1597--1603.
\newblock Available from:
  \url{http://www.ncbi.nlm.nih.gov/pmc/articles/PMC3183755/}.

\bibitem{rheo_crosslinkslip2}
Lieleg O, Schmoller KM, Claessens MMAE, Bausch AR.
\newblock Cytoskeletal Polymer Networks: Viscoelastic Properties are Determined
  by the Microscopic Interaction Potential of Cross-links.
\newblock Biophysical Journal. 2009 6;96(11):4725--4732.
\newblock Available from:
  \url{http://www.sciencedirect.com/science/article/pii/S0006349509007589}.

\bibitem{theo_crosslinkslip3}
Lieleg O, Bausch AR.
\newblock Cross-Linker Unbinding and Self-Similarity in Bundled Cytoskeletal
  Networks.
\newblock Phys Rev Lett. 2007 Oct;99:158105.
\newblock Available from:
  \url{http://link.aps.org/doi/10.1103/PhysRevLett.99.158105}.

\bibitem{rheo_crosslinksmatter}
Wachsstock DH, Schwarz WH, Pollard TD.
\newblock Cross-linker dynamics determine the mechanical properties of actin
  gels.
\newblock Biophysical Journal. 1994;66(3, Part 1):801 -- 809.
\newblock Available from:
  \url{http://www.sciencedirect.com/science/article/pii/S0006349594808562}.

\bibitem{rheo_crosslinkslip1}
Lieleg O, Claessens MMAE, Luan Y, Bausch AR.
\newblock Transient Binding and Dissipation in Cross-Linked Actin Networks.
\newblock Phys Rev Lett. 2008 Sep;101:108101.
\newblock Available from:
  \url{http://link.aps.org/doi/10.1103/PhysRevLett.101.108101}.

\bibitem{rheo_crosslinkslip3}
Yao NY, Becker DJ, Broedersz CP, Depken M, MacKintosh FC, Pollak MR, et~al.
\newblock Nonlinear Viscoelasticity of Actin Transiently Cross-linked with
  Mutant alpha-Actinin-4.
\newblock Journal of Molecular Biology. 2011;411(5):1062 -- 1071.
\newblock Available from:
  \url{http://www.sciencedirect.com/science/article/pii/S0022283611007376}.

\bibitem{rheo_nonaffine}
Liu J, Koenderink GH, Kasza KE, MacKintosh FC, Weitz DA.
\newblock Visualizing the Strain Field in Semiflexible Polymer Networks: Strain
  Fluctuations and Nonlinear Rheology of $F$-Actin Gels.
\newblock Phys Rev Lett. 2007 May;98:198304.
\newblock Available from:
  \url{http://link.aps.org/doi/10.1103/PhysRevLett.98.198304}.

\bibitem{Robin:2014aa}
Robin FB, McFadden WM, Yao B, Munro EM.
\newblock Single-molecule analysis of cell surface dynamics in Caenorhabditis
  elegans embryos.
\newblock Nat Meth. 2014 06;11(6):677--682.
\newblock Available from: \url{http://dx.doi.org/10.1038/nmeth.2928}.

\bibitem{Fritzsche:2013aa}
Fritzsche M, Lewalle A, Duke T, Kruse K, Charras G.
\newblock Analysis of turnover dynamics of the submembranous actin cortex.
\newblock Molecular Biology of the Cell. 2013 03;24(6):757--767.
\newblock Available from:
  \url{http://www.ncbi.nlm.nih.gov/pmc/articles/PMC3596247/}.

\bibitem{Fritzschee1501337}
Fritzsche M, Erlenk{\"a}mper C, Moeendarbary E, Charras G, Kruse K.
\newblock Actin kinetics shapes cortical network structure and mechanics.
\newblock Science Advances. 2016;2(4).
\newblock Available from:
  \url{http://advances.sciencemag.org/content/2/4/e1501337}.

\bibitem{Carlsson:2010aa}
Carlsson AE.
\newblock Actin Dynamics: From Nanoscale to Microscale.
\newblock Annual review of biophysics. 2010 06;39:91--110.
\newblock Available from:
  \url{http://www.ncbi.nlm.nih.gov/pmc/articles/PMC2967719/}.

\bibitem{Lai:2008aa}
Lai FP, Szczodrak M, Block J, Faix J, Breitsprecher D, Mannherz HG, et~al.
\newblock Arp2/3 complex interactions and actin network turnover in
  lamellipodia.
\newblock The EMBO Journal. 2008 04;27(7):982--992.
\newblock Available from:
  \url{http://www.ncbi.nlm.nih.gov/pmc/articles/PMC2265112/}.

\bibitem{Van-Goor:2012aa}
Van~Goor D, Hyland C, Schaefer AW, Forscher P.
\newblock The Role of Actin Turnover in Retrograde Actin Network Flow in
  Neuronal Growth Cones.
\newblock PLoS ONE. 2012;7(2):e30959.
\newblock Available from:
  \url{http://www.ncbi.nlm.nih.gov/pmc/articles/PMC3281045/}.

\bibitem{1367-2630-14-3-033037}
Lenz M, Gardel ML, Dinner AR.
\newblock Requirements for contractility in disordered cytoskeletal bundles.
\newblock New Journal of Physics. 2012;14(3):033037.
\newblock Available from:
  \url{http://stacks.iop.org/1367-2630/14/i=3/a=033037}.

\bibitem{PhysRevX.4.041002}
Lenz M.
\newblock Geometrical Origins of Contractility in Disordered Actomyosin
  Networks.
\newblock Phys Rev X. 2014 Oct;4:041002.
\newblock Available from:
  \url{http://link.aps.org/doi/10.1103/PhysRevX.4.041002}.

\bibitem{rheo_2D1}
Murrell MP, Gardel ML.
\newblock F-actin buckling coordinates contractility and severing in a
  biomimetic actomyosin cortex.
\newblock Proceedings of the National Academy of Sciences.
  2012;109(51):20820--20825.
\newblock Available from:
  \url{http://www.pnas.org/content/109/51/20820.abstract}.

\bibitem{Alvarado:2013aa}
Alvarado J, Sheinman M, Sharma A, MacKintosh FC, Koenderink GH.
\newblock Molecular motors robustly drive active gels to a critically connected
  state.
\newblock Nat Phys. 2013 09;9(9):591--597.
\newblock Available from: \url{http://dx.doi.org/10.1038/nphys2715}.

\bibitem{Murrell15062014}
Murrell M, Gardel ML.
\newblock Actomyosin sliding is attenuated in contractile biomimetic cortices.
\newblock Molecular Biology of the Cell. 2014;25(12):1845--1853.
\newblock Available from:
  \url{http://www.molbiolcell.org/content/25/12/1845.abstract}.

\bibitem{Ennomani2016616}
Ennomani H, Letort G, Gu{\'e}rin C, Martiel JL, Cao W, N{\'e}d{\'e}lec F,
  et~al.
\newblock Architecture and Connectivity Govern Actin Network Contractility.
\newblock Current Biology. 2016;26(5):616 -- 626.
\newblock Available from:
  \url{http://www.sciencedirect.com/science/article/pii/S0960982216000543}.

\bibitem{Reymann1310}
Reymann AC, Boujemaa-Paterski R, Martiel JL, Gu{\'e}rin C, Cao W, Chin HF,
  et~al.
\newblock Actin Network Architecture Can Determine Myosin Motor Activity.
\newblock Science. 2012;336(6086):1310--1314.
\newblock Available from:
  \url{http://science.sciencemag.org/content/336/6086/1310}.

\bibitem{Ndlec:1997aa}
Nedelec FJ, Surrey T, Maggs AC, Leibler S.
\newblock Self-organization of microtubules and motors.
\newblock Nature. 1997 09;389(6648):305--308.
\newblock Available from: \url{http://dx.doi.org/10.1038/38532}.

\bibitem{Surrey1167}
Surrey T, N{\'e}d{\'e}lec F, Leibler S, Karsenti E.
\newblock Physical Properties Determining Self-Organization of Motors and
  Microtubules.
\newblock Science. 2001;292(5519):1167--1171.
\newblock Available from:
  \url{http://science.sciencemag.org/content/292/5519/1167}.

\bibitem{2015arXiv150706182H}
{Hiraiwa} T, {Salbreux} G.
\newblock {Role of turn-over in active stress generation in a filament
  network}.
\newblock ArXiv e-prints. 2015 Jul;.

\bibitem{Mak:2016aa}
Mak M, Zaman MH, Kamm RD, Kim T.
\newblock Interplay of active processes modulates tension and drives phase
  transition in self-renewing, motor-driven cytoskeletal networks.
\newblock Nat Commun. 2016 01;7.
\newblock Available from: \url{http://dx.doi.org/10.1038/ncomms10323}.

\bibitem{10.1371/journal.pone.0000696}
Zumdieck A, Kruse K, Bringmann H, Hyman AA, J{\"u}licher F.
\newblock Stress Generation and Filament Turnover during Actin Ring
  Constriction.
\newblock PLoS ONE. 2007 08;2(8):e696.
\newblock Available from:
  \url{http://dx.plos.org/10.1371/journal.pone.0000696}.

\bibitem{PhysRevLett.113.148102}
Dierkes K, Sumi A, Solon J, Salbreux G.
\newblock Spontaneous Oscillations of Elastic Contractile Materials with
  Turnover.
\newblock Phys Rev Lett. 2014 Oct;113:148102.
\newblock Available from:
  \url{http://link.aps.org/doi/10.1103/PhysRevLett.113.148102}.

\bibitem{theo_friction}
Vanossi A, Manini N, Urbakh M, Zapperi S, Tosatti E.
\newblock \textit{Colloquium} : Modeling friction: From nanoscale to mesoscale.
\newblock Rev Mod Phys. 2013 Apr;85:529--552.
\newblock Available from:
  \url{http://link.aps.org/doi/10.1103/RevModPhys.85.529}.

\bibitem{theo_frictionSam}
Spruijt E, Sprakel J, Lemmers M, Stuart MAC, van~der Gucht J.
\newblock Relaxation Dynamics at Different Time Scales in Electrostatic
  Complexes: Time-Salt Superposition.
\newblock Phys Rev Lett. 2010 Nov;105:208301.
\newblock Available from:
  \url{http://link.aps.org/doi/10.1103/PhysRevLett.105.208301}.

\bibitem{theo_molefric}
Filippov AE, Klafter J, Urbakh M.
\newblock Friction through Dynamical Formation and Rupture of Molecular Bonds.
\newblock Phys Rev Lett. 2004 Mar;92:135503.
\newblock Available from:
  \url{http://link.aps.org/doi/10.1103/PhysRevLett.92.135503}.

\bibitem{theo_frictionShila}
Banerjee S, Marchetti MC, M\"uller-Nedebock K.
\newblock Motor-driven dynamics of cytoskeletal filaments in motility assays.
\newblock Phys Rev E. 2011 Jul;84:011914.
\newblock Available from:
  \url{http://link.aps.org/doi/10.1103/PhysRevE.84.011914}.

\bibitem{salbreuxbphs}
Chugh P, Clark AG, Smith MB, Cassani DAD, Charras G, Salbreux G, et~al.
\newblock Nanoscale Organization of the Actomyosin Cortex during the Cell
  Cycle.
\newblock Biophysical Journal. 2016 2016/07/18;110(3):198a.
\newblock Available from: \url{http://dx.doi.org/10.1016/j.bpj.2015.11.1105}.

\bibitem{rheo_2D2}
Sanchez T, Chen DTN, DeCamp SJ, Heymann M, Dogic Z.
\newblock Spontaneous motion in hierarchically assembled active matter.
\newblock Nature. 2012 11;491(7424):431--434.
\newblock Available from: \url{http://dx.doi.org/10.1038/nature11591}.

\bibitem{theo_crosslinknonlinear}
Broedersz CP, Storm C, MacKintosh FC.
\newblock Effective-medium approach for stiff polymer networks with flexible
  cross-links.
\newblock Phys Rev E. 2009 Jun;79:061914.
\newblock Available from:
  \url{http://link.aps.org/doi/10.1103/PhysRevE.79.061914}.

\bibitem{theo_hlm}
Head DA, Levine AJ, MacKintosh FC.
\newblock Deformation of Cross-Linked Semiflexible Polymer Networks.
\newblock Phys Rev Lett. 2003 Sep;91:108102.
\newblock Available from:
  \url{http://link.aps.org/doi/10.1103/PhysRevLett.91.108102}.

\bibitem{theo_hlm2}
Wilhelm J, Frey E.
\newblock Elasticity of Stiff Polymer Networks.
\newblock Phys Rev Lett. 2003 Sep;91:108103.
\newblock Available from:
  \url{http://link.aps.org/doi/10.1103/PhysRevLett.91.108103}.

\bibitem{mol_fric}
Ward A, Hilitski F, Schwenger W, Welch D, Lau AWC, Vitelli V, et~al.
\newblock Solid friction between soft filaments.
\newblock Nat Mater. 2015 03;advance online publication:--.
\newblock Available from: \url{http://dx.doi.org/10.1038/nmat4222}.

\bibitem{theo_hydroish2}
Chandran PL, Mofrad MRK.
\newblock Averaged implicit hydrodynamic model of semiflexible filaments.
\newblock Phys Rev E. 2010 Mar;81:031920.
\newblock Available from:
  \url{http://link.aps.org/doi/10.1103/PhysRevE.81.031920}.

\bibitem{Unterberger2014}
Unterberger MJ, Holzapfel GA.
\newblock Advances in the mechanical modeling of filamentous actin and its
  cross-linked networks on multiple scales.
\newblock Biomechanics and Modeling in Mechanobiology. 2014;13(6):1155--1174.
\newblock Available from: \url{http://dx.doi.org/10.1007/s10237-014-0578-4}.

\bibitem{mccrum1997principles}
McCrum NG, Buckley CP, Bucknall CB.
\newblock Principles of Polymer Engineering.
\newblock Oxford science publications. Oxford University Press; 1997.
\newblock Available from: \url{https://books.google.com/books?id=EiqWQgAACAAJ}.

\bibitem{Kim2014526}
Kim T, Gardel ML, Munro E.
\newblock Determinants of Fluidlike Behavior and Effective Viscosity in
  Cross-Linked Actin Networks.
\newblock Biophysical Journal. 2014;106(3):526 -- 534.
\newblock Available from:
  \url{http://www.sciencedirect.com/science/article/pii/S0006349513058487}.

\bibitem{rheo_active}
Koenderink GH, Dogic Z, Nakamura F, Bendix PM, MacKintosh FC, Hartwig JH,
  et~al.
\newblock An active biopolymer network controlled by molecular motors.
\newblock Proceedings of the National Academy of Sciences of the United States
  of America. 2009 09;106(36):15192--15197.
\newblock Available from:
  \url{http://www.ncbi.nlm.nih.gov/pmc/articles/PMC2741227/}.

\bibitem{Gorfinkiel2011531}
Gorfinkiel N, Blanchard GB.
\newblock Dynamics of actomyosin contractile activity during epithelial
  morphogenesis.
\newblock Current Opinion in Cell Biology. 2011;23(5):531 -- 539.
\newblock Cell-to-cell contact and extracellular matrix.
\newblock Available from:
  \url{http://www.sciencedirect.com/science/article/pii/S0955067411000834}.

\bibitem{Theriot1991}
Theriot JA, Mitchison TJ.
\newblock Actin microfilament dynamics in locomoting cells.
\newblock Nature. 1991 Jul;352(6331):126--131.
\newblock Available from: \url{http://dx.doi.org/10.1038/352126a0}.

\bibitem{Murthy2016}
Murthy K, Wadsworth P.
\newblock Myosin-II-Dependent Localization and Dynamics of F-Actin during
  Cytokinesis.
\newblock Current Biology. 2016 2016/12/15;15(8):724--731.
\newblock Available from: \url{http://dx.doi.org/10.1016/j.cub.2005.02.055}.

\bibitem{Watanabe1083}
Watanabe N, Mitchison TJ.
\newblock Single-Molecule Speckle Analysis of Actin Filament Turnover in
  Lamellipodia.
\newblock Science. 2002;295(5557):1083--1086.
\newblock Available from:
  \url{http://science.sciencemag.org/content/295/5557/1083}.

\bibitem{Guha2016}
Guha M, Zhou M, Wang Yl.
\newblock Cortical Actin Turnover during Cytokinesis Requires Myosin II.
\newblock Current Biology. 2016 2016/12/15;15(8):732--736.
\newblock Available from: \url{http://dx.doi.org/10.1016/j.cub.2005.03.042}.

\bibitem{Fritzsche15032013}
Fritzsche M, Lewalle A, Duke T, Kruse K, Charras G.
\newblock Analysis of turnover dynamics of the submembranous actin cortex.
\newblock Molecular Biology of the Cell. 2013;24(6):757--767.
\newblock Available from:
  \url{http://www.molbiolcell.org/content/24/6/757.abstract}.

\end{thebibliography}

\section*{S1 Appendix}
\subsection*{A.1 Simulation and Analysis Code Available Online}
All of the simulation and analysis code for generating the figures in this paper is available online.  To find the source code please visit our Github repository at 

https://github.com/wmcfadden/activnet

\subsection*{A.2 Steady-state Approximation of Effective Viscosity}
\label{sec:eff_vic}
We begin with a calculation of a strain rate estimate of the effective viscosity for a network described by our model in the limit of highly rigid filaments.  We carry this out by assuming we have applied a constant stress along a transect of the network.  With moderate stresses, we assume the network reaches a steady state affine creep. In this situation, we would find that the stress in the network exactly balances the sum of the drag-like forces from cross-link slip.  So for any transect of length D, we have a force balance equation.

\begin{equation}
\mathbf{\sigma} = \frac{1}{D}\sum_{filaments}\: \sum_{crosslinks}\xi \cdot (\mathbf{v_i(x)}-\mathbf{v_j(x)})
\end{equation}

where $\mathbf{v_i(x)}-\mathbf{v_j(x)}$ is the difference between the velocity of a filament, $i$, and the velocity of the filament, $j$, to which it is attached at the cross-link location, $\mathbf{x}$. We can convert the sum over cross-links to an integral over the length using the average density of cross-links, $1/l_c$ and invoking the assumption of (linear order) affine strain rate, $\mathbf{v_i(x)}-\mathbf{v_j(x)}=\dot \gamma x$. This results in

\begin{multline}
\mathbf{\sigma} =  \frac{1}{D}\sum_{filaments}\:  \int_0^L \xi \cdot  \: (\mathbf{v_i(s)}-\mathbf{v_j(s)}) \:\frac{ds \cos \theta }{l_c} \\
= \sum_{filaments}\:  \frac{\xi \dot \gamma L}{l_c} \cos \theta \cdot (x_l + \frac{L}{2} \cos \theta)
\end{multline}

Here we have introduced the variables $x_l$, and $\theta$ to describe the leftmost endpoint and the angular orientation of a given filament respectively.  Next, to perform the sum over all filaments we convert this to an integral over all orientations and endpoints that intersect our line of stress. We assume for simplicity that filament stretch and filament alignment are negligible in this low strain approximation.  Therefore, the max distance for the leftmost endpoint is the length of a filament, L, and the maximum angle as a function of endpoint is $\arccos(x_l/L)$.  The linear density of endpoints is the constant $D/l_cL$ so our integrals can be rewritten as this density over $x_l$ and $\theta$ between our maximum and minimum allowed bounds.

\begin{equation}
\mathbf{\sigma} =  \frac{1}{D} \int_0^L dx_l \int_{-\arccos (\frac{x_l}{L})}^{\arccos (\frac{x_l}{L})}\pi d\theta \frac{\xi \dot \gamma L}{l_c} \cdot \frac{D}{Ll_c}\cdot (x_l \cos \theta + \frac{L}{2} cos^2\theta)
\end{equation}

Carrying out the integrals and correcting for dangling filament ends leaves us with a relation between stress and strain rate.

\begin{equation}
\mathbf{\sigma} = 4 \pi \left ( \frac{ L}{l_c}-1 \right)^2 \xi \dot \gamma 
\end{equation}

We recognize the constant of proportionality between stress and strain rate as a viscosity (in 2 dimensions).  Therefore, our approximation for the effective viscosity, $\eta_{c}$, at steady state creep in this low strain limit is

\begin{equation}
\label{lin_eqn}
\mathbf{\sigma} = 4 \pi \left ( \frac{ L}{l_c}-1 \right)^2 \xi
\end{equation}

\subsection*{A.3 Critical filament lifetime for steady state filament extension}
We seek to determine a critical filament lifetime, $\tau_{crit}$ , below which the density of filaments approaches a stable steady state under constant extensional strain. To this end, let $\rho$ be the filament density (i.e. number of filaments per unit area). We consider a simple coarse grained model for how $\rho$ changes as a function of filament assembly $k_{ass}$, filament disassembly $k_{diss}$, $\rho$ and strain thinning $\dot{\gamma}\rho$. Using $\rho_0 = \frac{k_{ass}}{k_{diss}}$, $\tau_r=\frac{1}{k_{diss}}$, and $\dot{\gamma}=\frac{\sigma}{\eta_c}$.

\begin{equation}
\label{drho_1}
\frac{d\rho}{dt}=\frac{1}{\tau_r}\left ( \rho_0 - \rho - \frac{\sigma \tau_r}{\eta_c(\rho)} \rho\right )
\end{equation}

where $\eta_c = \eta_c(\rho)$ on the right hand side reflects the dependence of effective viscosity on network density.  The strength of this dependence determines whether there exists a stable steady state, representing continuous flow.  Using $\eta_c(\rho)\sim \xi \left ( \frac{L}{l_c(\rho)} -1 \right )^2$ from above (ignoring the numerical prefactor) and $\rho \sim \frac{2}{L l_c(\rho)}$, we obtain:

\begin{equation}
\label{drho_2}
\frac{d\rho}{dt}=\frac{1}{\tau_r}\left ( \rho_0 - \rho - \frac{\sigma \tau_r}{\xi(\rho L^2/2 -1)^2}\rho\right )
\end{equation}

\begin{figure}[h!]
	\centering
	\includegraphics[width=\hsize]{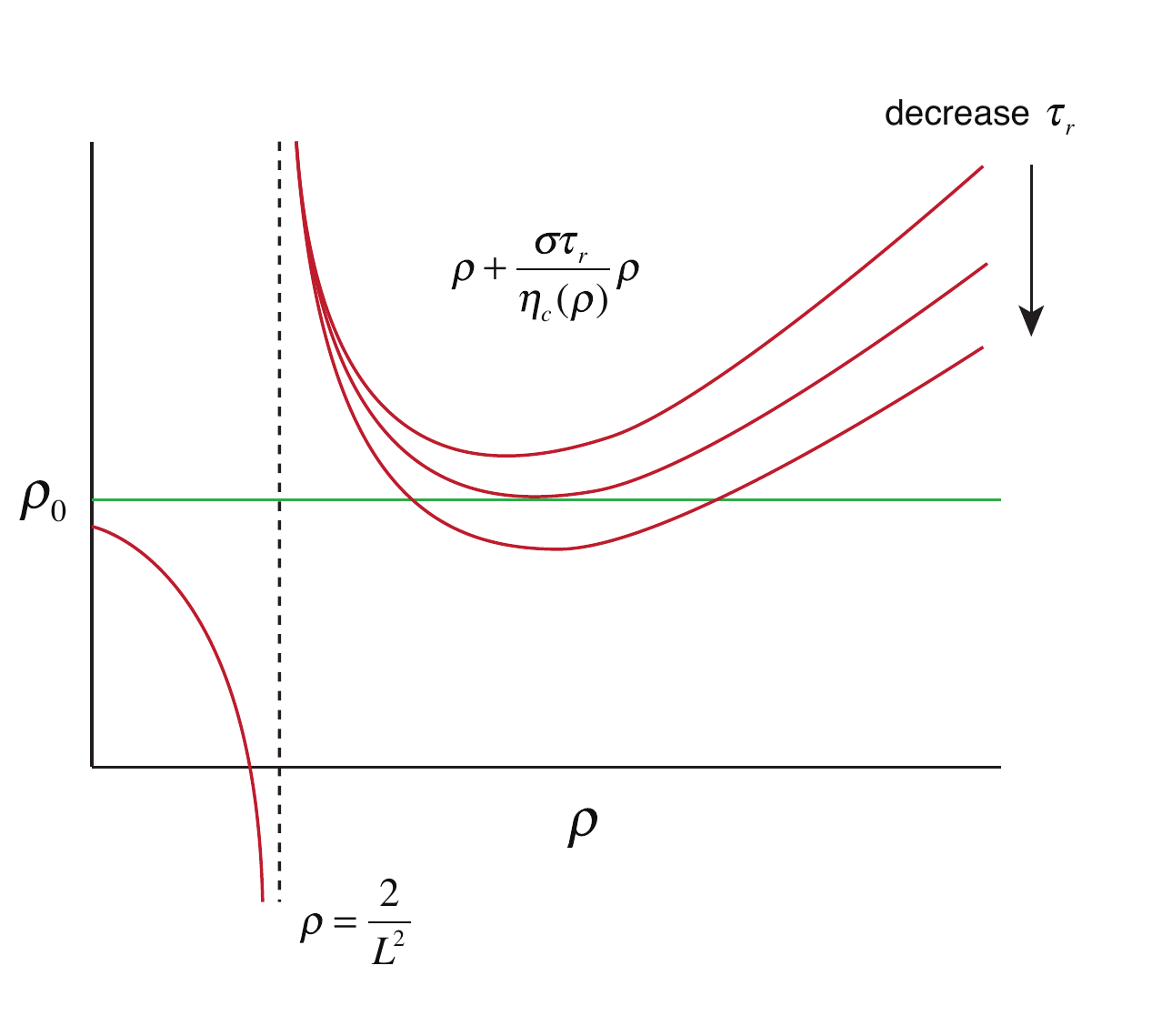}
	\caption*{ \textbf{Fig. A.1 } Flux balance analysis of network density. Qualitative plots of $\rho+\frac{\sigma \tau_r}{\eta_c(\rho)}\rho$ (red curves) vs $\rho_0$ (green line) for different values of $\tau_r$.  For sufficiently large $\tau_r$, there are no crossings.  For $\tau_r < \tau_{crit}$, there are two crossings:  The rightmost crossing represents a stable steady state.  }
\end{figure}

Fig. A.1 sketches the positive ($\rho_0$) and negative ($\rho+\frac{\sigma \tau_r}{\eta_c(\rho)}\rho$) contributions to the right hand side of Equation 6 for different values of $\tau_r$. For sufficiently large $\tau_r$, there is no stable state, i.e. strain thinning will occur.  However, as $\tau_r$ decreases below a critical value $\tau_{crit}$, a stable steady state appears.  Note that when $\tau_r = \tau_{crit}$, $\rho+\frac{\sigma\tau_r}{\eta_c(\rho)}\rho$ passes through a minimum value $\rho_0$ at $\rho=\rho^*$.  Accordingly, to determine $\tau_{crit}$, we solve:

\begin{equation}
\label{drho_3}
0 = \frac{d}{d\rho}\left( \rho + \frac{\sigma\tau_r}{\eta_c(\rho)} \rho\right ) = 1 - \frac{\sigma\tau_r}{\xi (\rho L^2/2-1)^3}
\end{equation}

From this, with some algebra, we infer that

\begin{equation}
\label{drho_4}
\rho^* = \frac{2}{L^2}\left ( 1 + \left( \frac{\sigma\tau_r}{\xi}\right )^{1/3} \right )
\end{equation}

and 

\begin{equation}
\label{drho_5}
\frac{\sigma\tau_r}{\eta_c(\rho^*)} =  \left( \frac{\sigma\tau_r}{\xi}\right )^{1/3} 
\end{equation}

We seek a value for $\tau_r=\tau_{crit}$ at which

\begin{equation}
\label{drho_6}
\rho^* + \frac{\sigma\tau_{crit}}{\eta_c(\rho^*)}\rho^* =  \rho_0
\end{equation}

Substituting from above, and using $\rho_0=\frac{2}{L l_c}$, we have:

\begin{equation}
\label{drho_7}
\frac{2}{L^2}\left ( 1 + \left( \frac{\sigma\tau_{crit}}{\xi}\right )^{1/3}  \right )
\left ( 1 + \left( \frac{\sigma\tau_{crit}}{\xi}\right )^{1/3}  \right )
= \frac{2}{L l_c}
\end{equation}

Finally, rearranging terms, we obtain

\begin{equation}
\label{drho_8}
\tau_{crit}=\frac{\xi}{\sigma}\left( \sqrt{\frac{L}{l_c}}-1\right )^3
\end{equation}

\begin{table}[h]
	\begin{adjustwidth}{-2.25in}{0in}
		\centering
		\caption{Parameter values sampled for individual figures}
		\label{table:para2}
		\begin{tabular}{|c|ccccccc|}
			\hline
			Parameter     & \textbf{Figure 3}          & \textbf{Figure 4}       & \textbf{Figure S3a,b}    & \textbf{Figure S3c,d}     & \textbf{Figure 7}    & \textbf{Figure 9}          \\ \hline
			$L$           & $1,3,5,7,10$      & $3$             & $3,5$         & $3,5$                 & $5$         & $3,5,8$            \\ 
			$l_c$         & $0.2,0.3,0.5,0.8$ & $0.3,0.5$       & $0.3$       & $0.15,0.2,0.3,0.4$ & $0.2,0.3$        & $0.15,0.2,0.3,0.4$ \\ 
			$\mu_e/\mu_c$ & $100$             & $100$           & $3-300$     & $100$     & $100$           & $100$              \\ 
			$\mu_c$       & $0.01$            & $0.01$          & $0.01-0.3$  & $0.001-0.03$       & $0.01$      & $0.01$             \\ 
			$\xi$         & $0.1,1$          & $0.05,0.1,1$      & $0.01,0.1,1$  & $0.1,1$  & $0.1,1,3.3$   & $0.1,1$           \\ 
			$\upsilon$    & ~                 & ~               & $0.1,0.3,1$ & $0.1,1$   & $0.1,1,3$    & $0.1$              \\ 
			$\phi$        & ~                 & ~               & $0.25$      & $0.5$     & $0.25,0.75$    & $0.25$             \\ 
			$\tau_r$      & ~                 & $0.1-10^4$      & ~           & ~         & $0.01-10^3$  & $0.01-10^3$              \\ 
			$\sigma$      & $0.0002-0.01$     & $0.00003-0.005$ & ~           & ~         & ~            & ~                          \\
			\hline
		\end{tabular}
	\end{adjustwidth}
\end{table}

\begin{figure}[h!]
	\centering
	\includegraphics[width=\hsize]{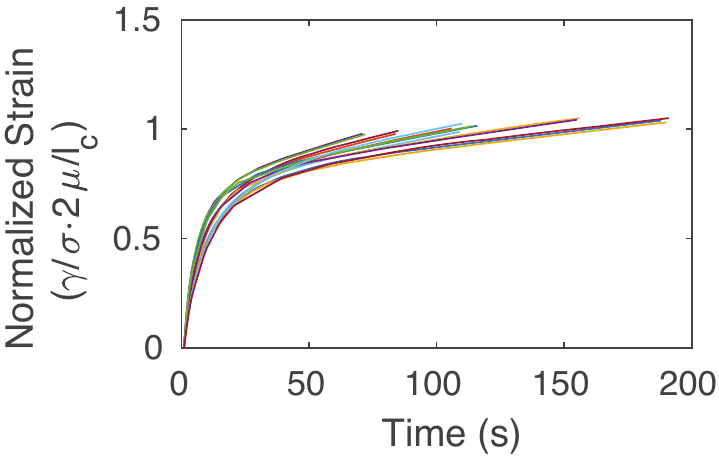}
	\caption*{\label{fig:passive_supp0}\textbf{S1 Fig.}  Fast viscoelastic response to extensional stress. Plots of normalized strain vs time during the elastic phase of deformation in passive networks under extensional stress.  Measured strain is normalized by the equilibrium strain predicted for a network of elastic filaments without crosslink slip $\gamma_{eq} = \sigma/G_0 = \sigma/(2\mu/l_c)$.  }
\end{figure}

\begin{figure}[h!]
	\centering
	\includegraphics[width=\hsize]{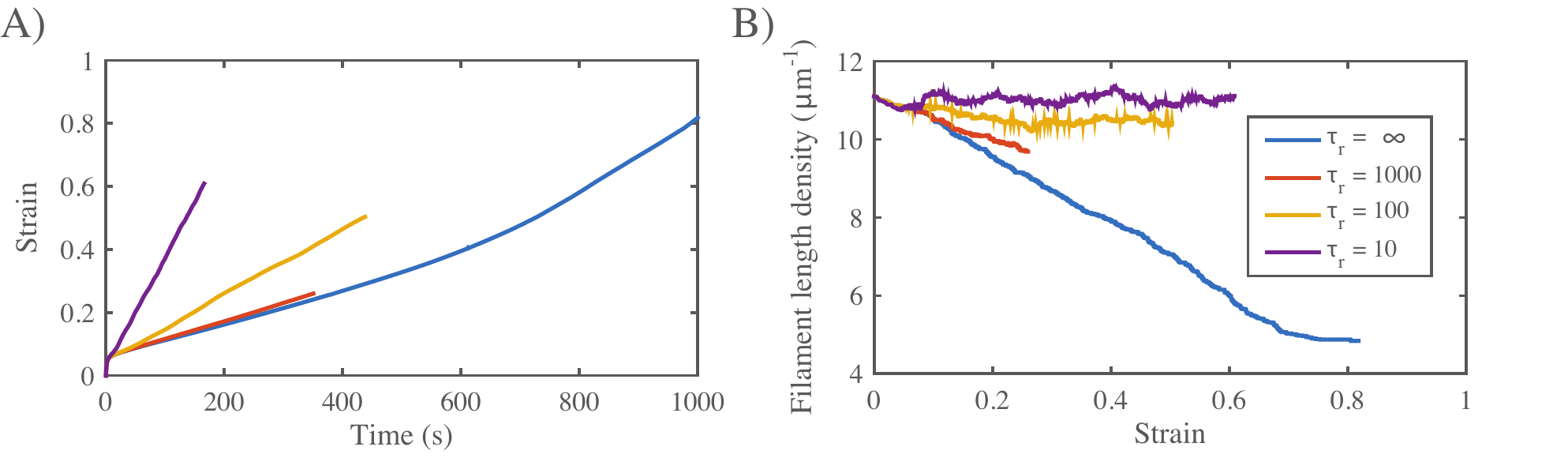}
	\caption*{\label{fig:thinning0} \textbf{S2 Fig.} Filament turnover rescues strain thinning.  \textbf{A)} Plots of strain vs time for different turnover times (see inset in (B)). Note the increase in strain rates with decreasing turnover time. \textbf{B)} Plots of filament density vs strain for different turnover times $\tau_r$.  For intermediate $\tau_r$, simulations predict progressive strain thinning, but at a lower rate than in the complete absence of recycling. For higher $\tau_r$, densities approach steady state values at longer times.  }
\end{figure}

\begin{figure}[h!]
	\centering
	\includegraphics[width=\hsize]{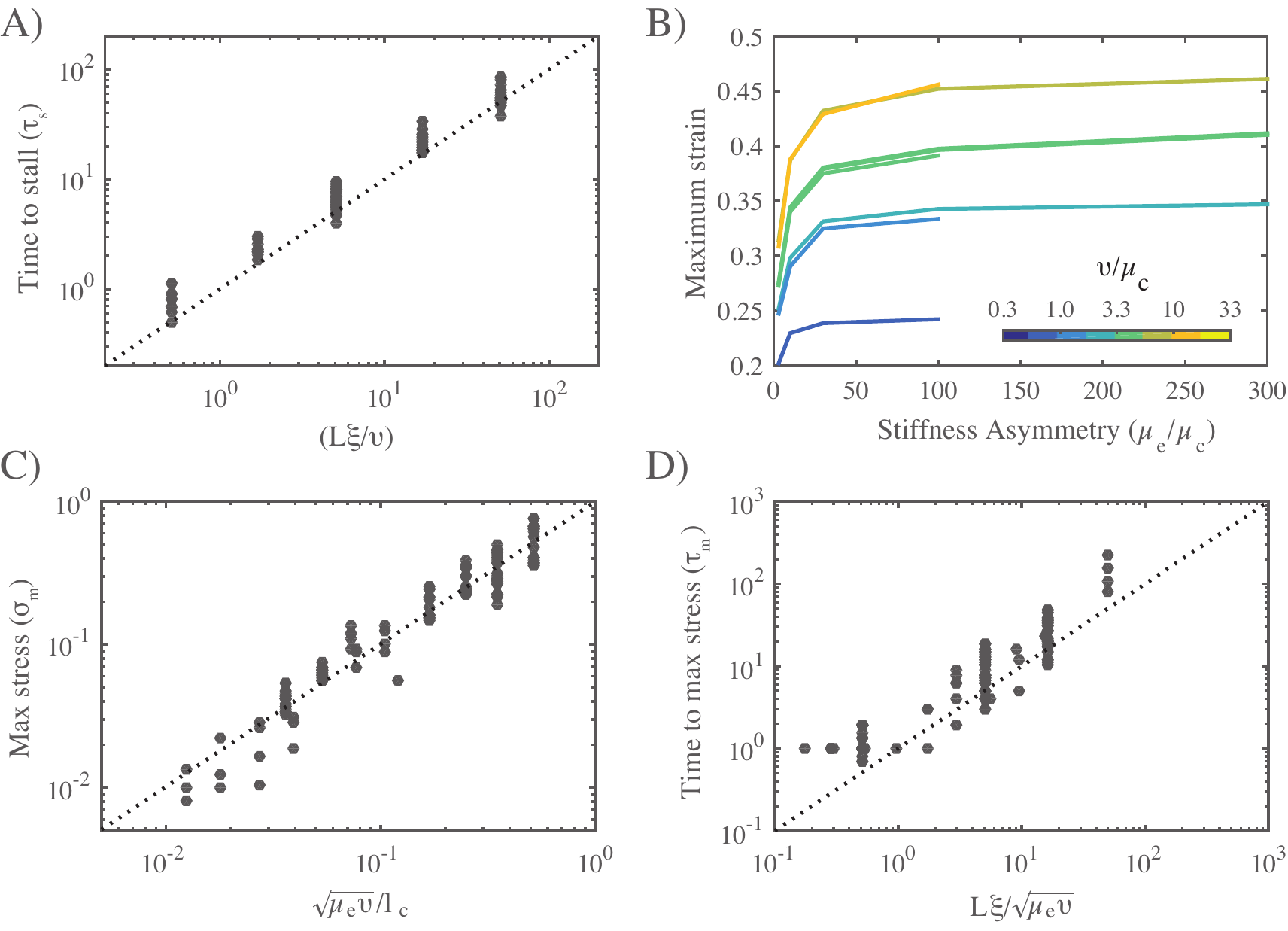}
	\caption*{\label{fig:active_supp0} \textbf{S3 Fig.} Mechanical properties of active networks.  \textbf{A)}  Time for freely contracting networks to reach maximum strain, $\tau_s$, scales with $L\xi/\upsilon$.  \textbf{B)} Free contraction requires asymmetric filament compliance, and total network strain increases with the applied myosin force $\upsilon$. Note that the maximum contraction approaches an asymptotic limit as the stiffness asymmetry approaches a ratio of $\sim 100$.   \textbf{C)}  Maximum stress achieved during isometric contraction, $\sigma_m$, scales approximately with $\sqrt{\mu_e\upsilon}/l_c$.  \textbf{D)} Time to reach max stress during isometric contraction scales approximately with $L\xi/\sqrt{\mu_e\upsilon}$. Scalings for $\tau_s$, $\sigma_m$ and $\tau_m$ were determined empirically by trial and error, guided by dimensional analysis. }
\end{figure}

\begin{figure}[h!]
	\centering
	\includegraphics[width=\hsize]{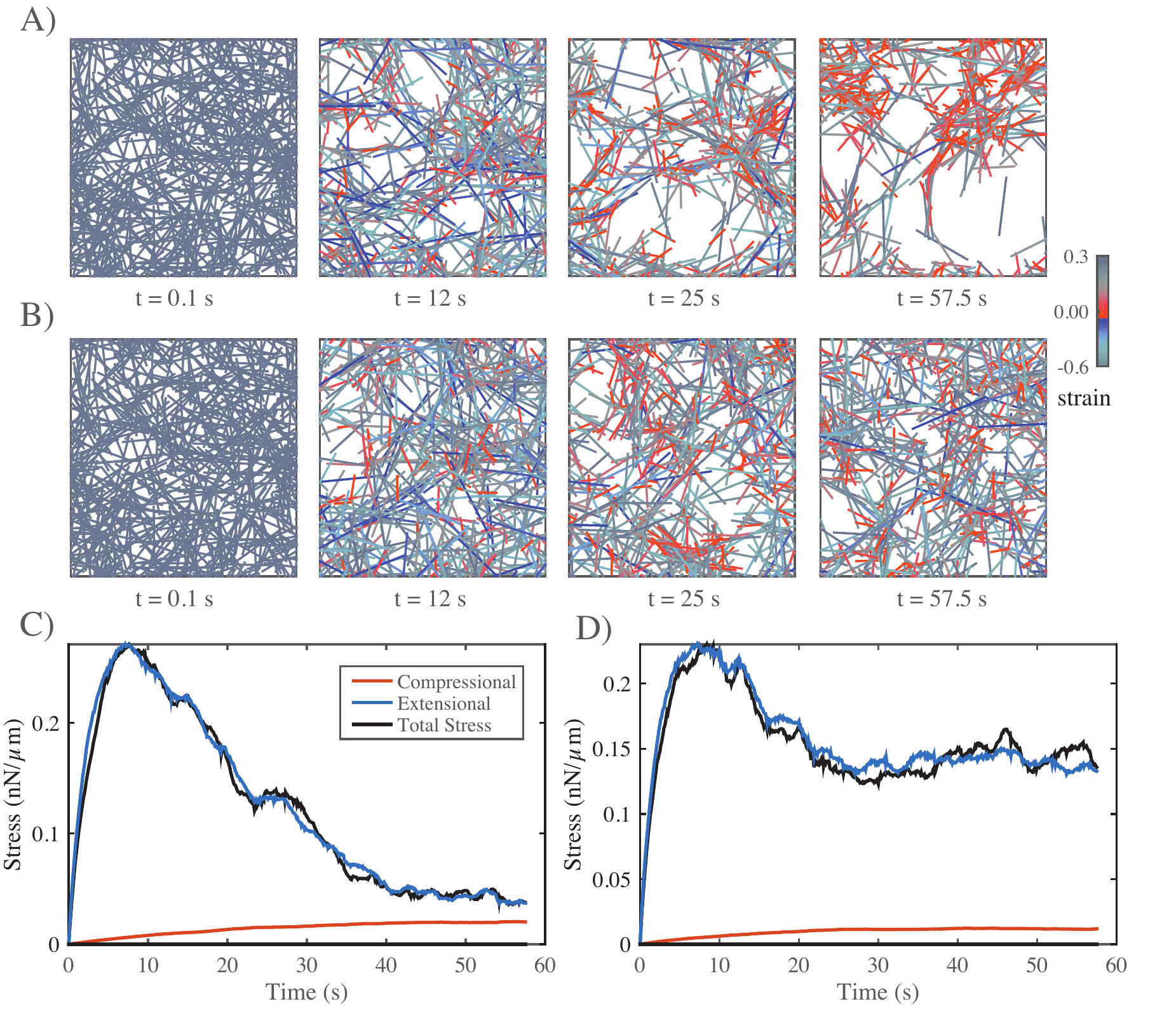}
	\caption*{\label{fig:active_tear0} \textbf{S4 Fig.} Filament turnover prevents tearing of active networks.  \textbf{A)}  An active network undergoing large scale deformations due to active filament rearrangements.  \textbf{B)}  The same network as in (A) but with a shorter filament turnover time.  \textbf{C)}  Plots of internal stress vs time for the network in (A).  \textbf{D)}  Plots of internal stress vs time for the network in (B).  }
\end{figure}

\begin{figure}[h!]
	\centering
	\includegraphics[width=\hsize]{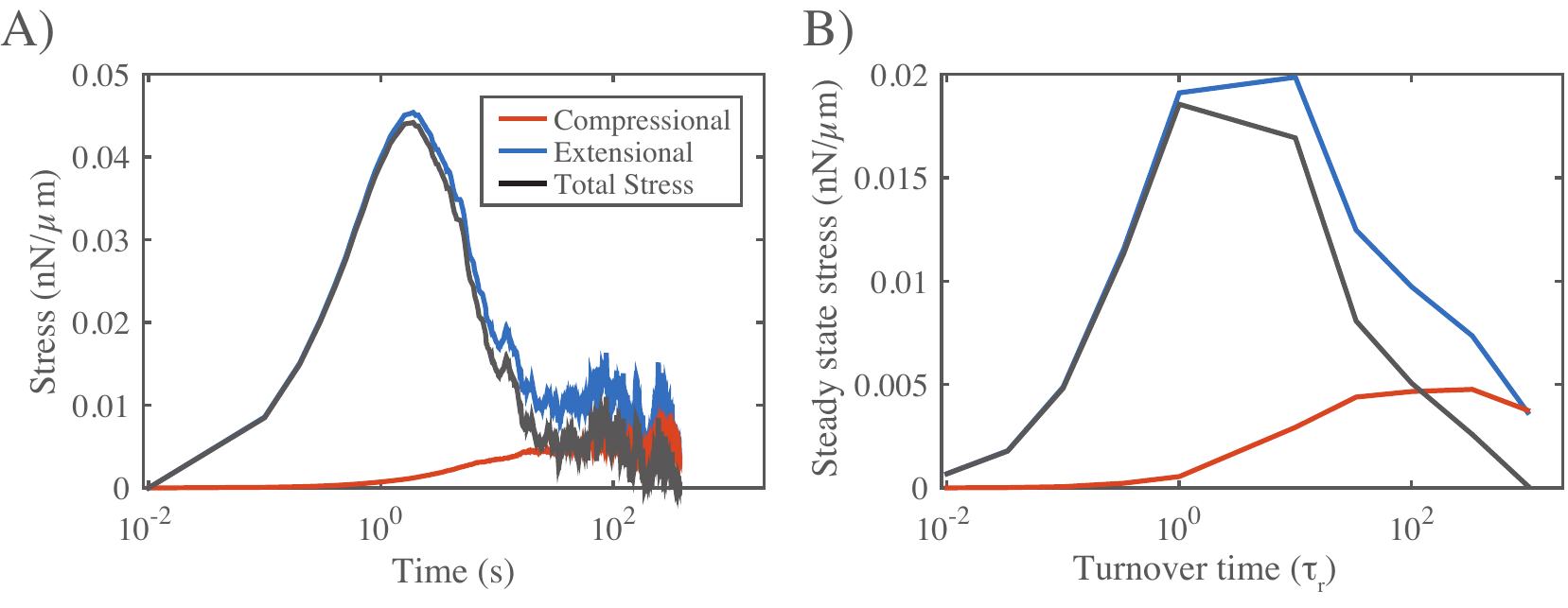}
	\caption*{\label{fig:recycle_supp0} \textbf{S5 Fig.}  Bimodal dependence on turnover time matches bimodal buildup and dissipation of stress in the absence of turnover.  \textbf{A)}  Bimodal buildup of stress in a network with very slow turnover ($\tau_r = 1000s$).  \textbf{B)}  Steady state stress for networks with same parameters as in (Aa), but  for a range of filament turnover times.   }
\end{figure}

\begin{figure}[h!]
	\centering
	\includegraphics[width=\hsize]{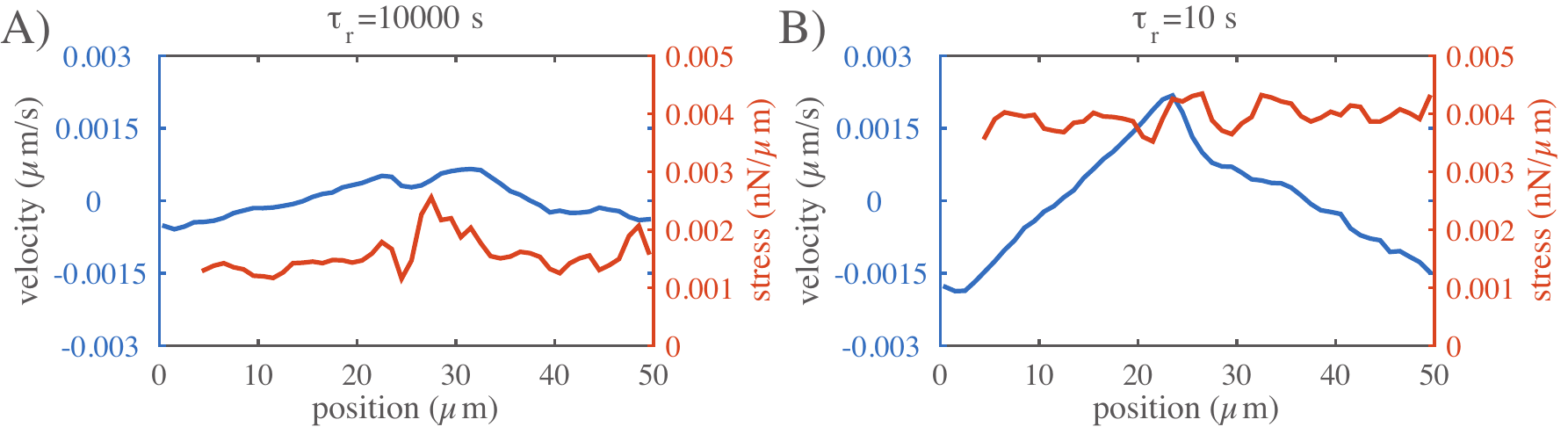}
	\caption*{\label{fig:combo_prof0} \textbf{S6 Fig.} Dynamics of steady state flow. Plots of stress and strain vs position for networks in which motor activity is limited to the right-half domain and filament turnover time is either  \textbf{A)} $\tau_r = 10000$ or  \textbf{B)} $\tau_r = 10 s$.  Blue indicates velocity while orange represents total stress, measured as described in the main text. }
\end{figure}
\end{document}